%% file: spie_speckle_lifetime_jmilli.tex
\documentclass[]{spie}  

\newcommand\Stot{$S_{\text{total}}$}
\newcommand\Ssub{$S_{\text{sub,k}}$}

\usepackage{amsmath,amsfonts,amssymb}
\usepackage{graphicx}
\usepackage[colorlinks=true, allcolors=blue]{hyperref}

\input{macros_spie}

\title{Speckle lifetime in XAO coronagraphic images: temporal evolution of SPHERE coronagraphic images}

\author[a]{J. Milli}
\author[b]{T. Banas}
\author[c]{D. Mouillet}
\author[d]{D. Mawet}
\author[a]{J. H. Girard}
\author[e]{A. Vigan}
\author[f]{A. Boccaletti}
\author[b]{M. Kasper}
\author[a]{Z. Wahhaj}
\author[c]{A.M. Lagrange}
\author[c]{J.-L. Beuzit}
\author[h]{T. Fusco}
\author[h]{J.-F. Sauvage}
\author[f]{R. Galicher}

\affil[a]{European Southern Observatory (ESO), Alonso de C\'ordova 3107, Vitacura, Casilla 19001, Santiago, Chile}
\affil[b]{ESO, Karl-Schwarzschild-Stra{\ss}e 2, 85748 Garching, Germany}
\affil[c]{Universit\'e Grenoble Alpes, IPAG, F-38000 Grenoble, France }
\affil[d]{Department of Astronomy, California Institute of Technology, 1200 E. California Blvd, MC 249-17, Pasadena, CA 91125 USA}
\affil[e]{Aix Marseille Universit\'e, CNRS, LAM (Laboratoire d'Astrophysique de Marseille) UMR 7326, 13388, Marseille, France}
\affil[f]{LESIA, Observatoire de Paris, CNRS, Universite Pierre et Marie Curie 6 and Universite Denis Diderot Paris 7, 5 place Jules Janssen, 92195 Meudon, France}
\affil[h]{ONERA, The French Aerospace Lab, BP72, 29 avenue de la Division Leclerc, 92322 Chatillon Cedex, France}

\authorinfo{Further author information: Julien Milli: E-mail: jmilli@eso.org	}

\begin{document} 
\maketitle

\begin{abstract}

The major source of noise in high-contrast imaging is the presence of slowly evolving speckles that do not average with time. The temporal stability of the point-spread-function (PSF) is therefore critical to reach a high contrast with extreme adaptive optics (xAO) instruments. Understanding on which timescales the PSF evolves and what are the critical parameters driving the speckle variability allow to design an optimal observing strategy and data reduction technique to calibrate instrumental aberrations and reveal faint astrophysical sources. We have obtained a series of 52 min, AO-corrected, coronagraphically occulted, high-cadence (1.6Hz), H-band images of the star HR\, 3484 with the SPHERE (Spectro-Polarimeter High-contrast Exoplanet REsearch \cite{Beuzit2008}) instrument on the VLT. This is a unique data set from an xAO instrument to study its stability on timescales as short as one second and as long as several tens of minutes. We find different temporal regimes of decorrelation. We show that residuals from the atmospheric turbulence induce a fast, partial decorrelation of the PSF over a few seconds, before a transition to a regime with a linear decorrelation with time, at a rate of several tens parts per million per second (ppm/s). We analyze the spatial dependence of this decorrelation, within the well-corrected radius of the adaptive optics system and show that the linear decorrelation is faster at short separations. Last, we investigate the influence of the distance to the meridian on the decorrelation. 

\end{abstract}

\keywords{Extreme Adaptive Optics,Speckles, High Contrast}

\section{INTRODUCTION}
\label{sec:intro}  

Direct imaging of exoplanets and circumstellar environments requires to probe very close separations, within a fraction of an arcsecond from the host star, and to supress the light from the star \cite{Oppenheimer2003}. A new generation of instruments have entered into operations in the last years, known as extreme adaptive optics (xAO) instruments, to push further down the achievable contrast, with the primary goal to better characterize the population of young gas giants orbiting beyond a few astronomical units of their host stars \cite{Beuzit2008,Macintosh2014,Close2012,Guyon2010,Dekany2013,Esposito2011}. These instruments involve a high-performance adaptive optics system with a dense deformable mirror and sensitive wavefront sensors to reach high Strehl ratios. They use in conjunction advanced coronagraphs to reject the starlight \cite{Mawet2012}. However, residual starlight still contaminates the area of interest. It comes from both the diffracted light of the telescope pupil and the wavefront errors due to uncorrected atmospheric perturbations or imperfect optics within the light path. The diffraction pattern can be attenuated by specific diffraction control techniques and is stable in time for pupil-stabilized observations. However the second source of residual starlight is more difficult to address. Wavefront errors create speckles in the image plane that mimic point sources and degrade the contrast \cite{Racine1999}. These speckles vary temporally on different timescales, from a few milliseconds for non-corrected atmospheric perturbations up to several minutes, hours or even days for quasi-static speckles slowly variable with the temperature, the telescope flexures or the movement of the optical elements in the light beam such as the derotator or the atmospheric dispersion corrector. In addition they can interfere with the static pattern to create pinned speckles \cite{Bloemhof2001}. Eventually, those quasi-static speckles represent the major source source of noise at short separations because they are correlated in time and do not average out during an observing sequence \cite{Hinkley2007}, unlike other sources of noise \cite{Macintosh2005}. 

To calibrate them, various observation strategies have been designed, based on differential imaging. One of the most successful is known as Angular Differential Imaging \cite{Marois2006} (ADI), and relies on the differential motion between the pupil and the sky to disentangle speckles from an astrophysical signal. A reference image is constructed as a linear combination of frames recorded at different parallactic angles from the science frame. Because this reference frame is built from images taken at different times, this method is intrisically subject to the variability of the speckles and will work best for a highly stable temporal sequence. We therefore aim here at analysing empirically how coronagraphic images evolve with time in the well-corrected region of an xAO instrument, based on on-sky images. This complements the analysis already made base on laboratory measurements\cite{Martinez2012,Martinez2013}.

\section{INSTRUMENTAL SETUP}

\subsection{Measurements}
\label{sec_measurements}
 
The observations were carried out on March 3, 2015 between 03:53 and 04:35UT on the bright star HR\,3484 of magnitude $R=3.6$ and $H=2.32$.
 They used the pupil-tracking mode of VLT / SPHERE, to keep the aberrations as stable as possible. To reach a high contrast, we made use of the coronagraphic combination N\_ALC\_YJH\_S corresponding to an apodizer, a Lyot mask of diameter 185\,mas and an undersized Lyot stop to block the starlight rejected off the mask and to cover the telescope spiders.
 
\begin{figure} [ht]
 \begin{center}
 \includegraphics[height=5cm]{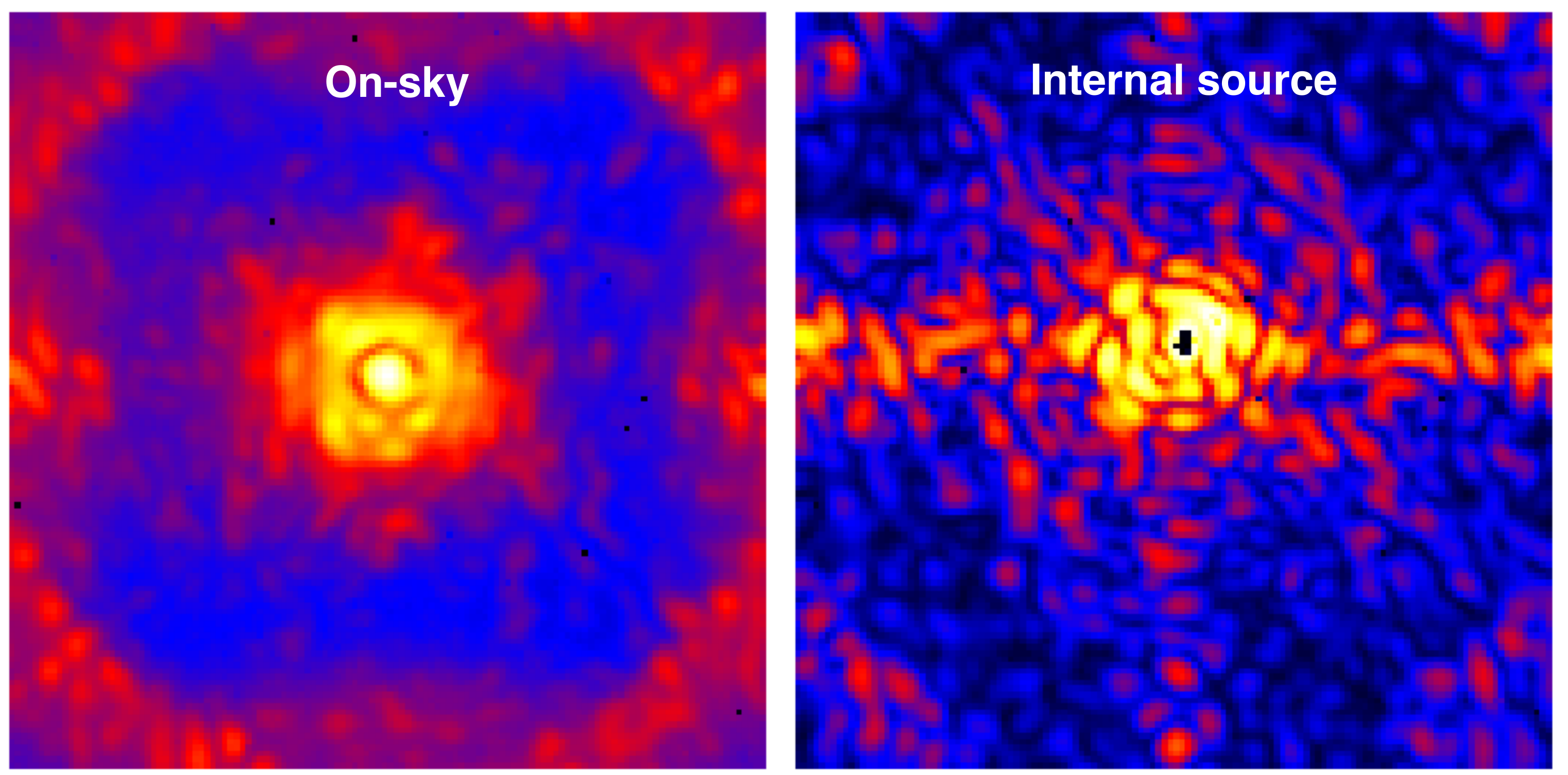}
 \end{center}
 \caption[example] 
{ \label{fig_coro_median_image} 
Median of the sequence of coronagraphic images obtained on-sky (left image) and with the internal source (right image). The color scale is logarithmic, the image is 1.57 \arcsec{} in the side. A video of both sequences is available online at \url{http://dx.doi.org/10.1117/12.2231703} (in the supplemental content section).}
\end{figure}

 We used the IRDIS subsystem \cite{Dohlen2008} in dual-band imaging \cite{Langlois2010}. The IRDIS imager splits the incoming light in two channels where the dual-band filters H2 and H3 were inserted ($\lambda_{H2}=1.588$\micron, $\Delta\lambda_{H2}=53.1$nm, $\lambda_{H3}=1.667$\micron, $\Delta\lambda_{H3}=55.6$nm). IRDIS provides a 11\arcsec$\times$11\arcsec{} field of view with a pixel scale of 12.25 mas. The star was observed after meridian passage, during 50 minutes under good and stable atmospheric conditions. The seeing was between 0.3\arcsec{} and 0.5\arcsec, the coherence time derived from DIMM measurements \cite{Sarazin2002} was between 3 and 4ms. The real time controller of the SPHERE AO system indicated a wind speed of 3 to 4m/s and derived a Strehl ratio of 92\% to 93\%. The spatial filter upstream the Shack-Hartmann wavefront sensor was set to its smallest aperture to get the least aliasing. In order to  obtain a high frame rate, the detector was windowed to read only 128 rows (out of a total of 1048), which enabled to use a detector integration time (DIT) of 0.1s. 5000 frames were recorded in this setup for a total of 3087s. This corresponds to a parallactic angle variation of $15^\circ$. We rejected the first 50 frames for which the AO loop had not reached a stationary regime. Because of the time to read the detector, the actual frame rate is one image per 617ms or 1.62Hz. The median frame of this sequence is shown in Fig. \ref{fig_coro_median_image} left.
 This deep coronagraphic sequence was followed by an acquisition of 500 frames on an empty sky region for measurement of the background, and by a point-spread function (hereafter PSF) measurement out of the coronagraphic mask with the neutral density filter ND\_2.
  
 To compare these on-sky measurements with data obtained without disturbance by the atmosphere, we recorded an additional sequence with the same instrumental setup but with the internal lamp and the instrument shutter closed, on November 24, 2015 between 01:08 and 02:17 UT. The sequence consists of 5740 frames with the AO loop closed. To be closer to on-sky conditions, the derotator was turned on to simulate pupil-tracking of a star at a declination of $-30^\circ$ above Paranal observed at an hour angle between -20\,min and +40\,min. The detector was read at the same speed in the same windowed region as for the on-sky observations. The median image of the sequence is shown in Fig. \ref{fig_coro_median_image} right. The only difference with on-sky observations in the instrument after the optical fiber injecting the light source is the atmospheric dispersion corrector, which is turned off while using the internal lamp.
 
 \begin{figure} [ht]
 \begin{center}
 \includegraphics[height=5cm]{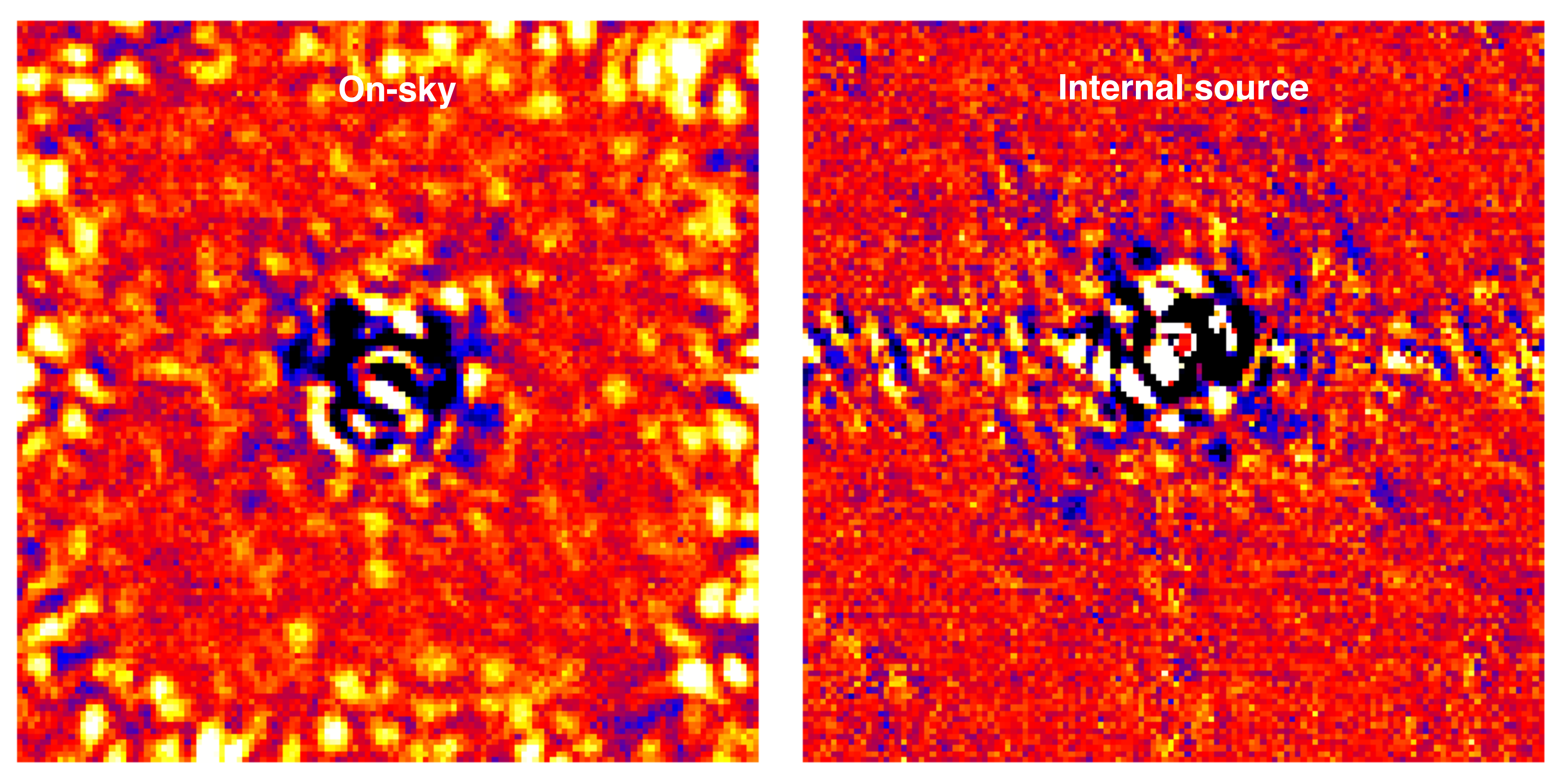}
 \end{center}
 \caption[example] 
{ \label{fig_coro_median_subtracted_image} 
Illustration of one of the images after subtraction of the median of the temporal sequence to remove the static part of the coronagraphic image. The color scale is linear, the image is 1.57\arcsec{} in the side.}
\end{figure} 

\subsection{Methodology}
\label{sec_methodo}

\begin{description}

\item[Definition] For each pair of coronagraphic images $(t_i,t_j)$, we computed the correlation coefficient $\rho(t_i,t_j)$ as follow:
\begin{align}
\label{eq_def_corr_coeff}
\rho(t_i,t_j)=\frac{\sum\limits_{x \in S} \left[ I(x,t_i) -  \overline{I_{t_i}} \right] \left[ I(x,t_j) -  \overline{I_{t_j}}\right] }  {\sigma_{t_i}\sigma_{t_j}} 
\end{align}

where $\overline{I_{t_{i}}}$ and $\sigma_{t_{i}}$ are the mean and standard deviation of the region \textsl{S} for the frame taken at time $t_{i}$. This is also known as the Pearson's correlation coefficient. The normalisation by $\sigma_{t_i} \sigma_{t_j}$ ensures a unit value for $t_i=t_j$. One of the challenge of this analysis is get a Signal to Noise ratio (S/N) large enough to allow to draw meaningfull conclusions. This was done by chosing regions \textsl{S} encompassing a sufficient number of pixels and by resorting to temporal binning in order to average the coeffficient correlations, as detailed below.
In Appendix \ref{app_contrast}) we show that such a definition of the correlation is directly related to the contrast obtained after subtracting one image of the pair from the other. 

\item[Global approach over \Stot]

In the first place, we computed the correlation coefficients $\rho(t_i,t_j)$ between each pair of images in a single region \Stot, encompassing all the pixels shown in Fig. \ref{fig_coro_median_image}, exluding those within 8px ($\sim98$ mas) of the central source, where the light is blocked by the coronagraphic mask. These coefficients can be displayed in the form of a symmetric matrix in Fig. \ref{fig_corr_coefficients_matrix} (left panel).

 \begin{figure} [ht]
 \begin{center}
  \includegraphics[height=5cm]{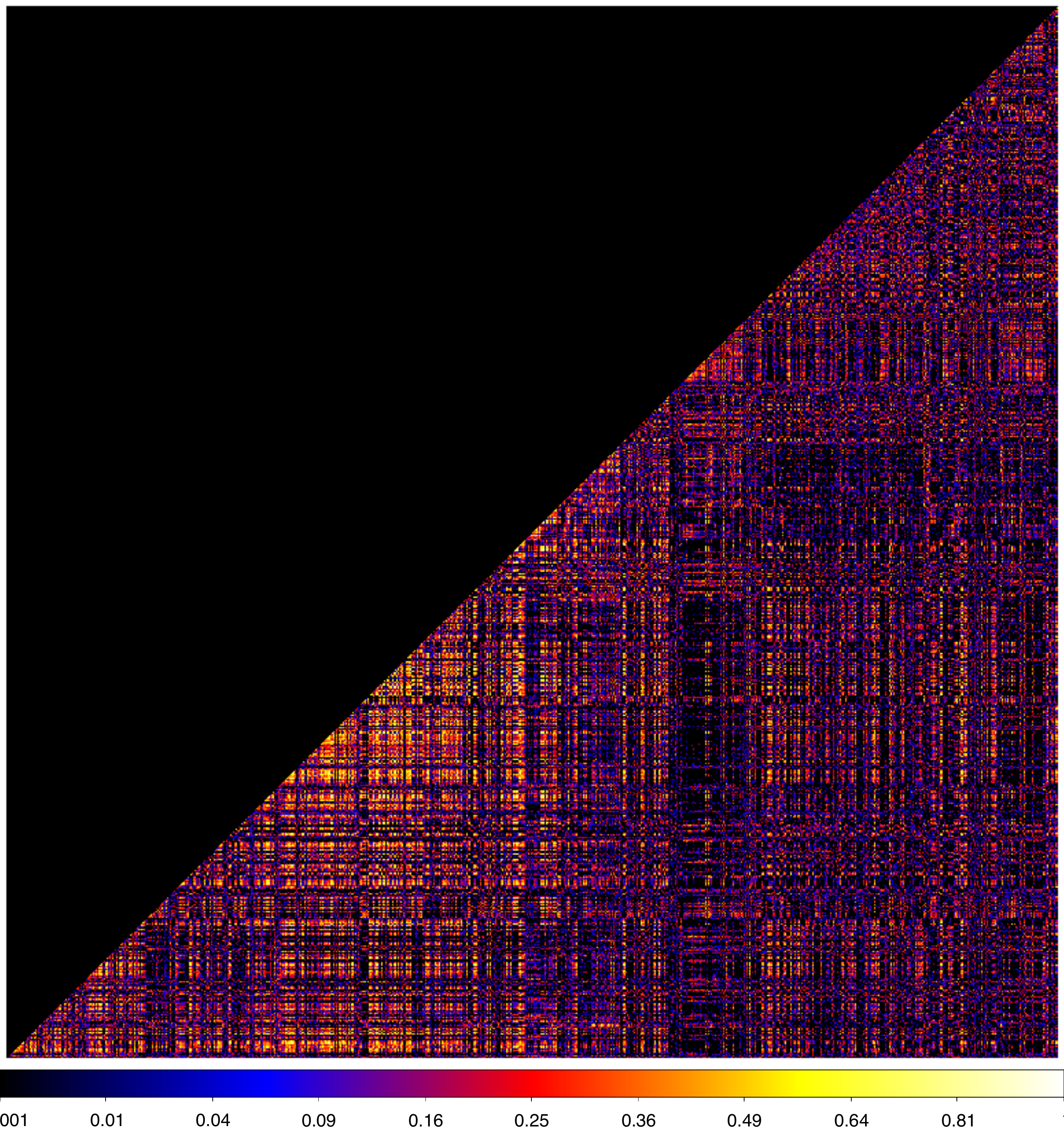}
 \includegraphics[height=5cm]{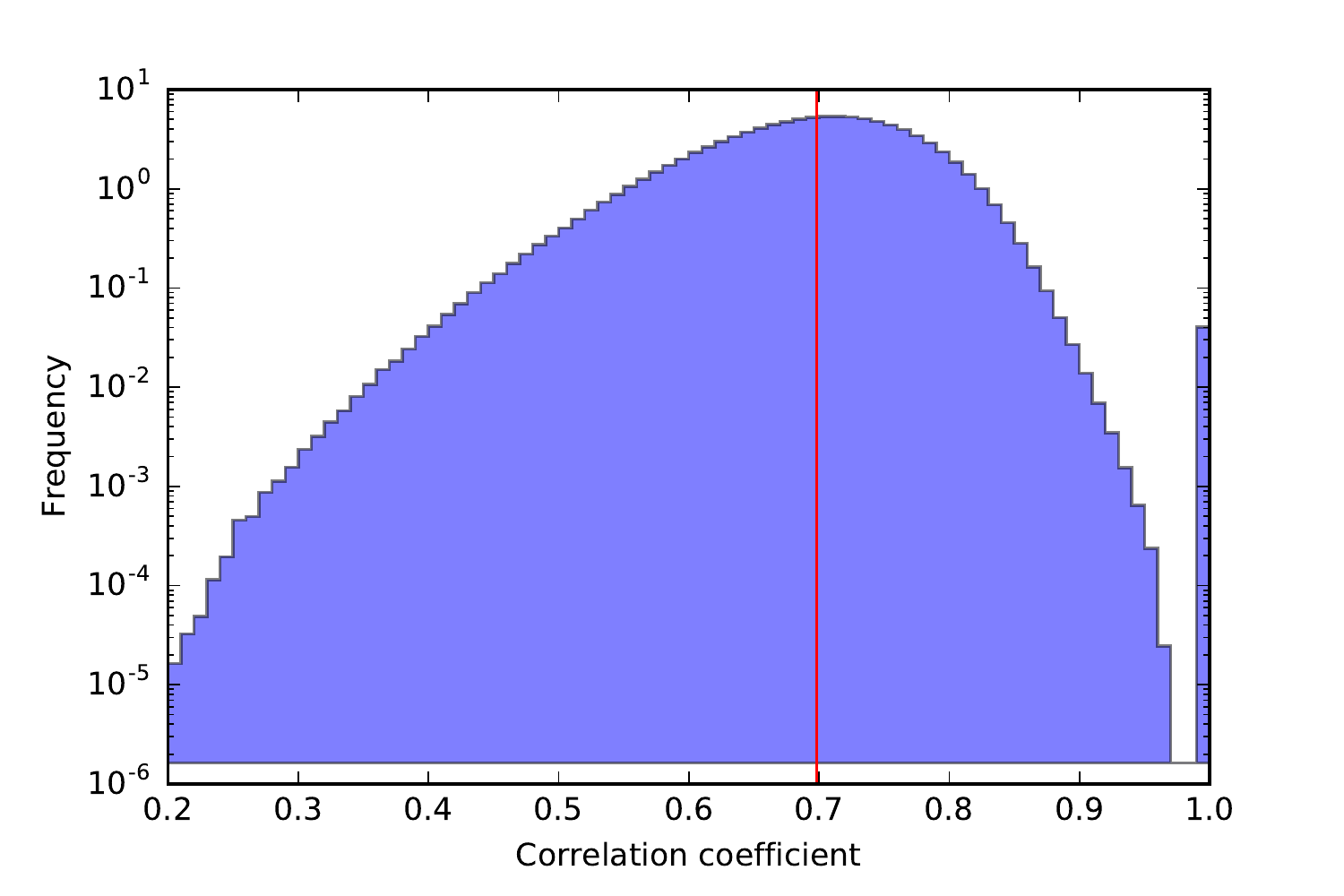}
 \end{center}
 \caption { \label{fig_corr_coefficients_matrix} 
Left: Visualisation of the correlation coefficients computed between each pair of images $(t_i,t_j)$ in the form of the squared matrix $\rho(t_i,t_j)$ of dimension $4950^2$. The matrix is symmetric, therefore only the $\sim1.2\times10^7$ elements below the diagonal are shown here ($j<i$). When one moves towards the lower right corner, the time ellapsed between the 2 images $t_j-t_i$ increases. The dark lines (horizontal or vertical) correspond to frames poorly correlated with the others. The colour scale is non-linear (square-root) from 0 (black) to 1 (white). These coefficients are computed in the whole region  \Stot. Right: Histogram of the correlation coefficients, normalized to an integral of one, the red vertical line is the median (0.7).}
\end{figure} 

All correlation coefficients span between 0.2 and 1, with a median of 0.70, as shown in the histogram of Fig. \ref{fig_corr_coefficients_matrix} (right). Still, to retrieve a significant trend from this histogram, additional processing and signal extraction is needed and this is the goal of the temporal binning explained below.



\item[Temporal binning]

To get a one-dimensional evolution of the correlation coefficients with time, we eventually averaged all coefficients calculated beween pairs separated by an equal time step. This enables to ease the interpretation by having a one-dimensional curve instead of a matrix and more importantly to get a higher S/N on the correlation by averaging a large number of values for 4950 frames. There are 4948 pairs separated by the elementary time step of 0.1\,s and this number decreases linearily as time goes by. Visually, this means that the two-dimensional matrix of Fig. \ref{fig_corr_coefficients_matrix} (left and middle panels) was averaged along the diagonal direction to be turned into a one-dimensional array of correlation coefficients that can directly be plotted against the time ellapsed between the frames. This is shown in Fig. \ref{fig_corrcoeff_vs_time_linearfit}. 

   \begin{figure} [ht]
   \begin{center}
   \begin{tabular}{c} 
   \includegraphics[width=0.6\hsize]{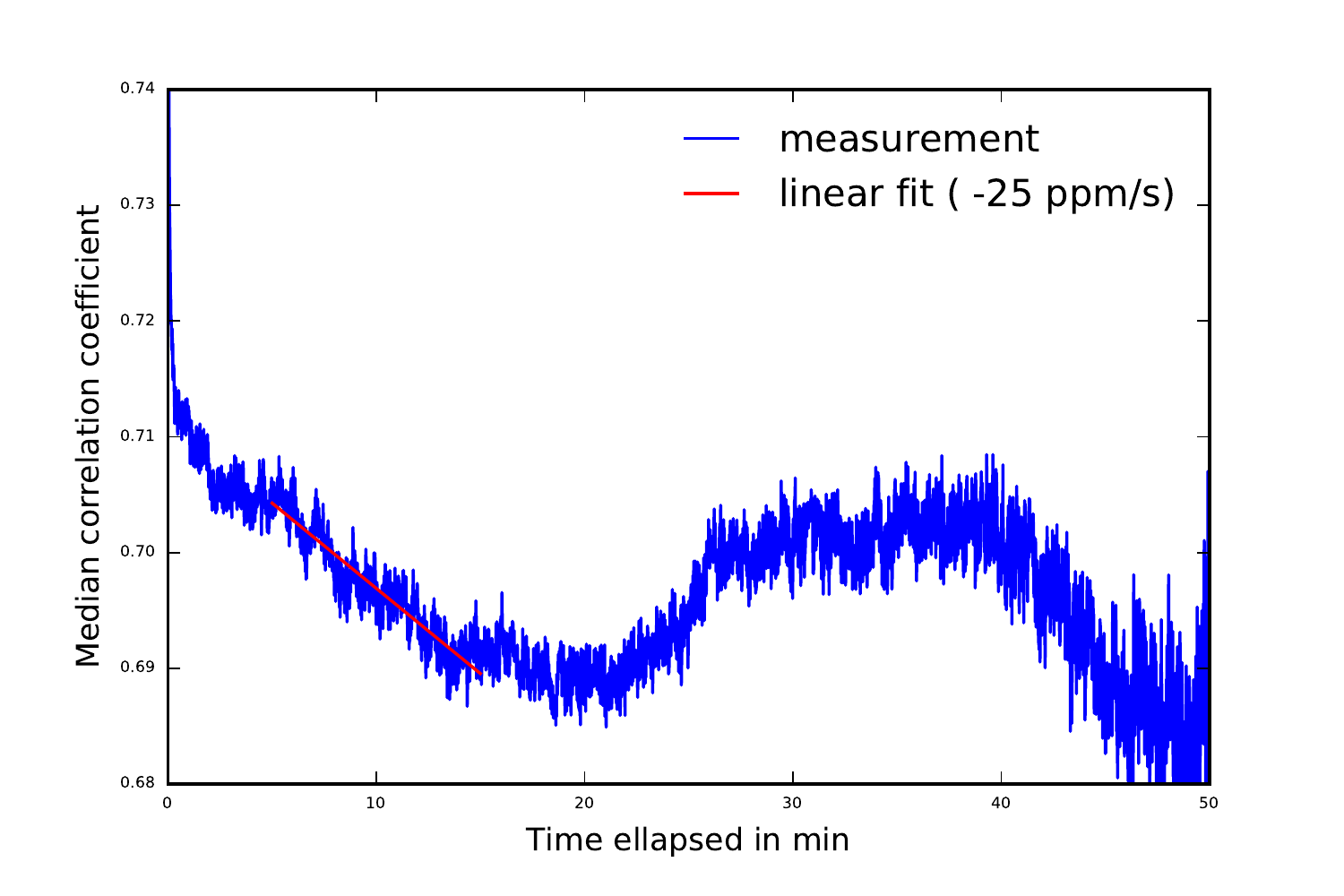}
	\end{tabular}
	\end{center}
   \caption   { \label{fig_corrcoeff_vs_time_linearfit} 
Evolution of the correlation coefficients calculated over the whole region \Stot, with time. A linear fit was performed between 5 and 15\,min.}
   \end{figure}

\item[Spatial analysis in subregions \Ssub{}]

In a second time, to get some spatial resolution for a zonal analysis, the image was divided in 12 by 12 squared regions of 10px wide (122mas or $3\lambda/D$), as shown in Fig. \ref{fig_grid_subregions}, written \Ssub{} with $1 \leq k \leq 148$,. For each subimage, we computed the correlation matrix $\rho(t_i,t_j)$ and averaged it to get a one-dimensional plot similar to that shown in Fig. \ref{fig_corrcoeff_vs_time_linearfit} for the whole image. These curves are shown in App. \ref{app_subregions}.

 \begin{figure} [ht]
 \begin{center}
 \includegraphics[height=5cm]{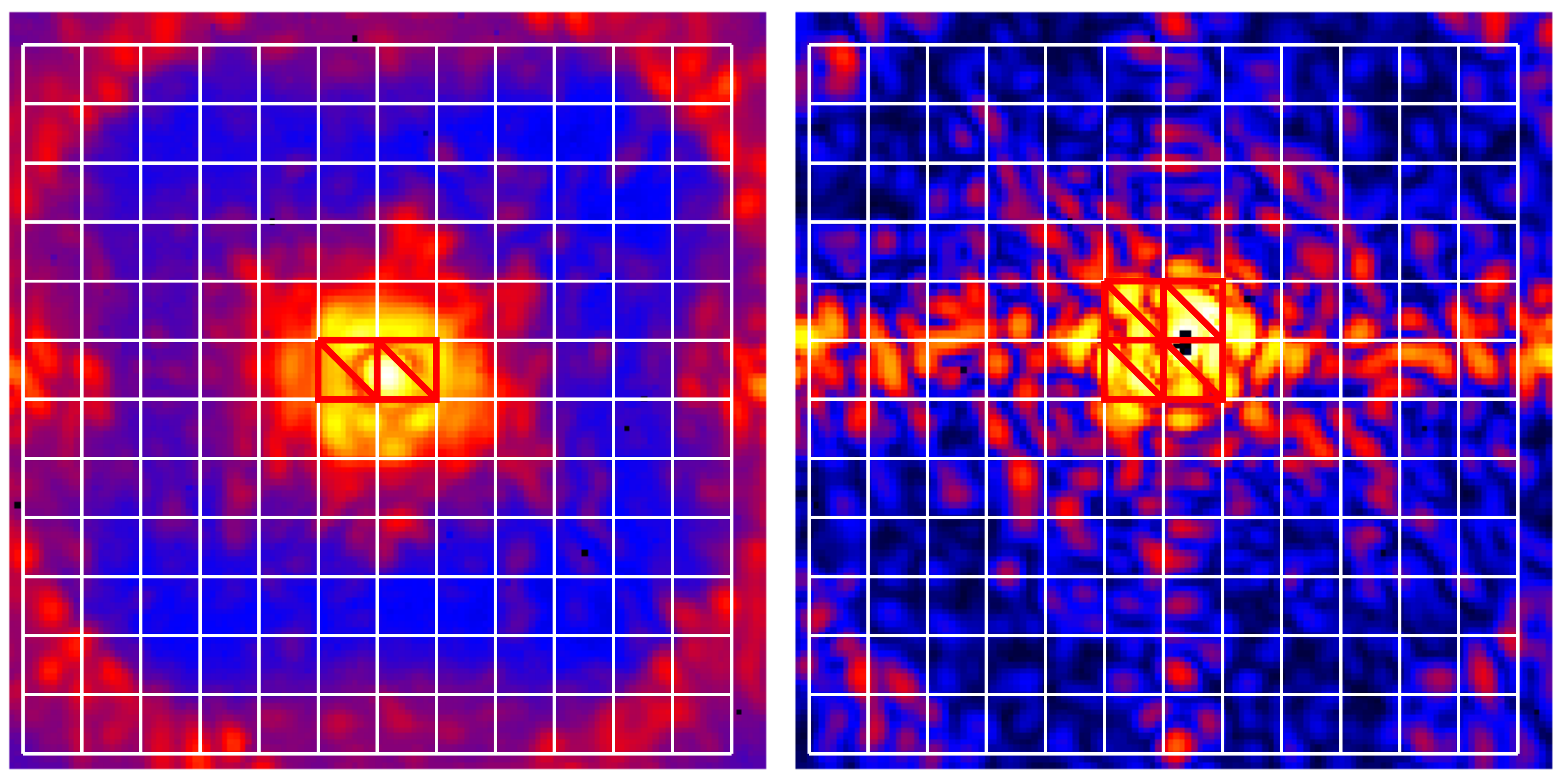}
 \end{center}
 \caption[example] 
{ \label{fig_grid_subregions} 
Same as Fig. \ref{fig_coro_median_image} with the grid of subregions used to calculate the correlation coefficients. Each square in the grid is 10px wide ($3\lambda/D$).}
\end{figure} 

\end{description}

\section{RESULTS}
\label{sec_results}

\subsection{Different regimes of decorrelation}
\label{sec_decorr_regime}

\begin{figure} [ht]
   \begin{center}
   \begin{tabular}{c} 
   \includegraphics[width=0.6\hsize]{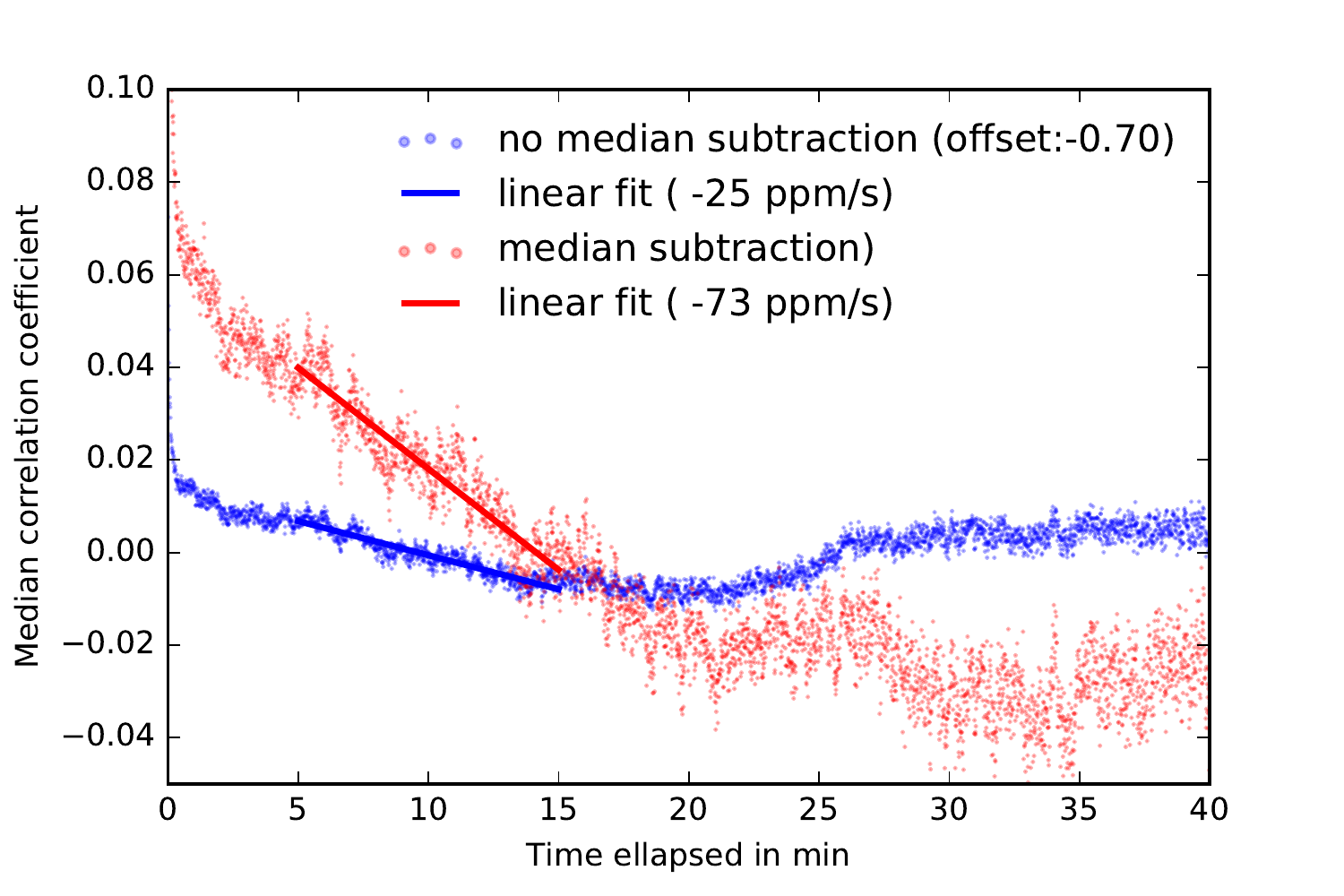}
	\end{tabular}
	\end{center}
   \caption[example] 
   { \label{fig_comparison_nomedsub_medsub} 
Comparison between the correlation coefficients computed in the original sequence (blue scatter points, identical as in Figure \ref{fig_corrcoeff_vs_time_linearfit}) and in the median-subtracted sequence (red scatter points). The blue curve was shifted down by 0.7 to be able to use the same vertical scale.}
   \end{figure}

The evolution of the correlation over time (Fig. \ref{fig_corrcoeff_vs_time_linearfit}) shows a linear regime of decorrelation over a timescale up to 20 min. The speed of decorrelation, computed as a linear fit between 5 and 15\,min reaches 25 parts per million per second (ppm/s) when the analysis is done in the whole region \Stot{}.
After 25\,min, the correlation increases again. To understand this behaviour, we carried out the same analysis after subtracting the median frame from the sequence in order to remove the static pattern of the coronagraphic image that creates the overall offset of the correlation by about 0.7. This is shown in Fig. \ref{fig_comparison_nomedsub_medsub} (red data points). In such a case, the correlation coefficients are much lower and the decrease in the correlation is much steeper (-73\,ppm/s) because this is a direct measure of the decorrelation of the residual, quasi-static, speckles. This confirms that the decorrelation observed in the non-median-subtracted sequence is real. In addition, we checked that we are not dominated by the detector or background noise (see App/\ref{app_sky} for a verification). The regime where the correlation increases again after 25min corresponds to a regime where the correlation computed on the median-subtracted sequence is negative. This means that the median-subtracted frames are anti-correlated. This is also visible in the individual frames of the sequence, such as that shown in Fig. \ref{fig_coro_median_subtracted_image} : because the speckle intensity varies over time, negative regions appear after subtraction of the temporal median of the sequence, creating an overall negative correlation coefficient. We refrained from interpretating this regime because the speckle pattern have evolved signifcantly on the globabl scale of the region \Stot.

\begin{figure} [ht]
   \begin{center}
   \begin{tabular}{c} 
   \includegraphics[width=0.6\hsize]{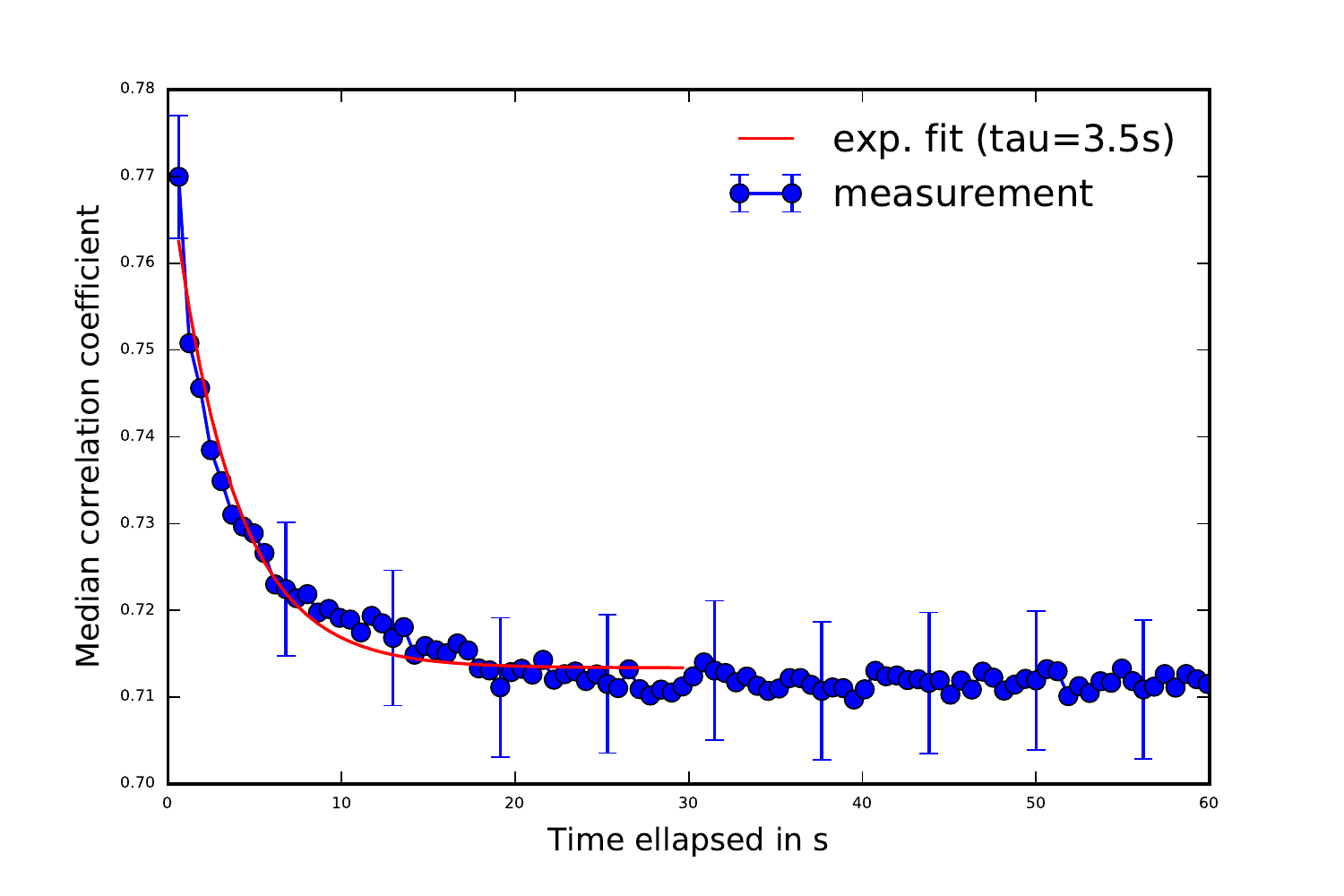}
	\end{tabular}
	\end{center}
   \caption[example] 
   { \label{fig_corrcoeff_vs_time_expfit} 
Zoom of Fig. \ref{fig_corrcoeff_vs_time_linearfit} over the first 60 seconds. A fit of an exponential profile was performed from 1 to 30\,s.}
   \end{figure}

Interestingly, there is a second decorrelation regime, corresponding to a much steeper decrease in the correlation coefficient over a smaller timescale of a few seconds. This is shown in Fig. \ref{fig_corrcoeff_vs_time_expfit}. An exponential decay was adjusted to the measurements in the form of the equation 

 \begin{align}
\label{eq_exp_decay}
\rho(t) = \Lambda e^{-t/\tau} + \rho_0
\end{align}

with $\tau = 3.5^{\pm0.2} $s, $\Lambda=0.059^{\pm0.002}$ and $\rho_0=0.713^{\pm0.0005}$.

\subsection{Zonal analysis: a faster linear decorrelation regime at small separations}
\label{sec_linear_fit_vs_sep}

\begin{figure} [ht]
   \begin{center}
   \begin{tabular}{c} 
   \includegraphics[width=0.6\hsize]{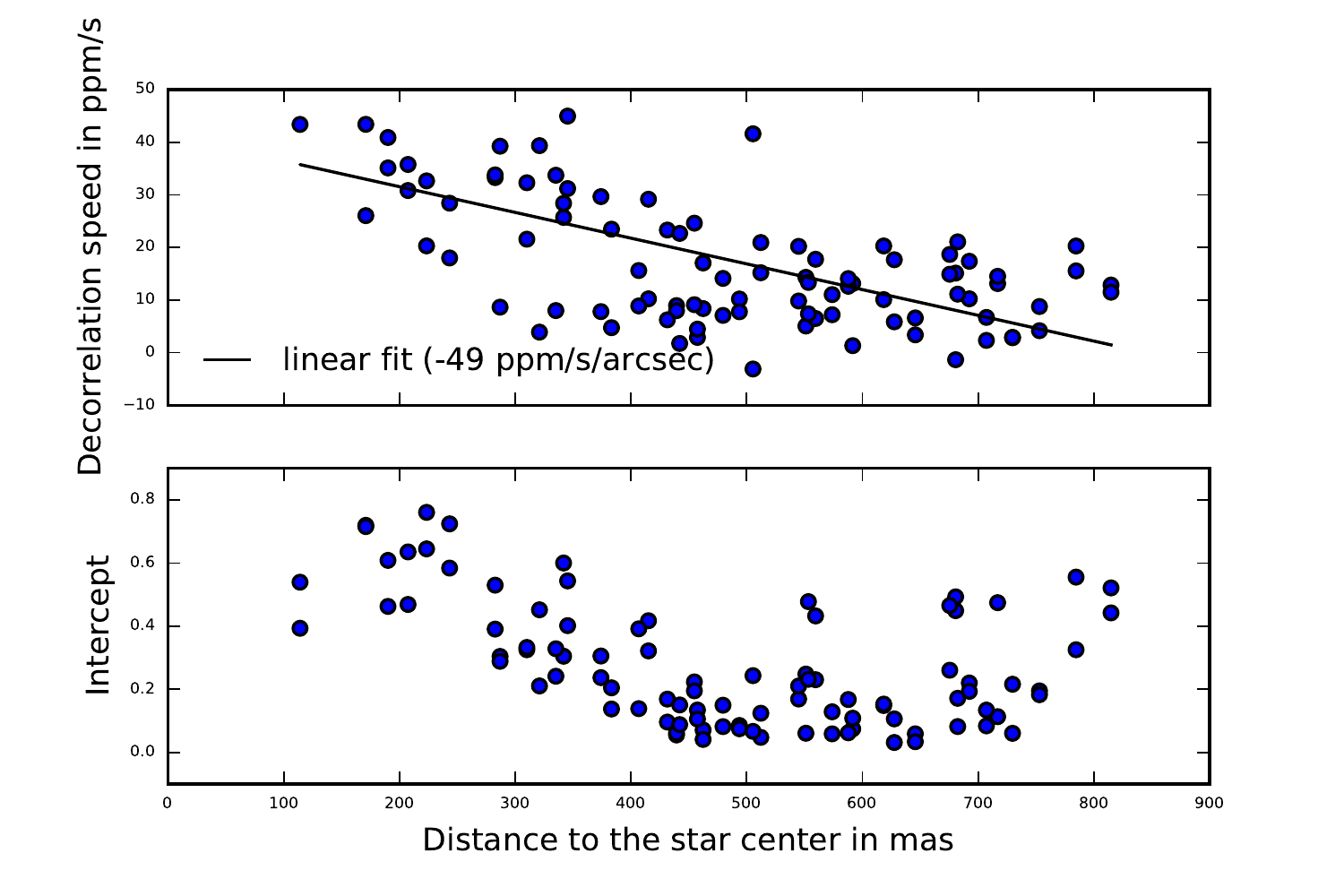}
	\end{tabular}
	\end{center}
   \caption[example] 
   { \label{fig_param_linear_fit_vs_sep} 
Parameters of the linear fit to the on-sky data. The top panel shows the slope of line, e.g. the decorrelation speed expressed in ppm/s and the bottom panel shows the intercept.}
   \end{figure}

The same analysis of the evolution of the correlation coefficient with time was made for the subregions \Ssub{} for $1 \leq k \leq 148$. We also see a linear decorrelation regime for ellapsed times between a few minutes and 30 minutes. An overall picture of the decorrelation plots is shown in Appendix \ref{app_subregions}.
In a second step, we performed a linear fit, as was done previously. We did not subtract the temporal median of each subregion. The parameters of the fit are plotted in Fig. \ref{fig_param_linear_fit_vs_sep} as a function of the separation between the star and the center of the subregion \Ssub. This plot shows that in the linear regime, the decorrelation is faster close to the optical axis, with a speed exceeding 30\,ppm/s for some subregions. We illustrate this trend by the black line in Fig. \ref{fig_param_linear_fit_vs_sep} (top), with a rate of decrease of $\sim50$ppm/s/\arcsec. Beyond 500\,mas no significant change is observerd and the speckles seem to evolve at a rate of $\sim10$ ppm/s. The intercept of the linear fit (Fig. \ref{fig_param_linear_fit_vs_sep} bottom) is overall following the profile of the coronagraphic image, which traces the fact that in the non-median subtracted sequence, the static part of the sequence has a major influence on the absolute value of the correlation coefficient (in case we subtract the temporal median of the sequence, the top curve shows a very similar trend but the bottom curve is flatter).

\subsection{Zonal analysis: a steep decorrelation independant of the separation}
\label{sec_exp_fit_vs_sep}

\begin{figure} [ht]
   \begin{center}
   \begin{tabular}{c} 
   \includegraphics[width=0.6\hsize]{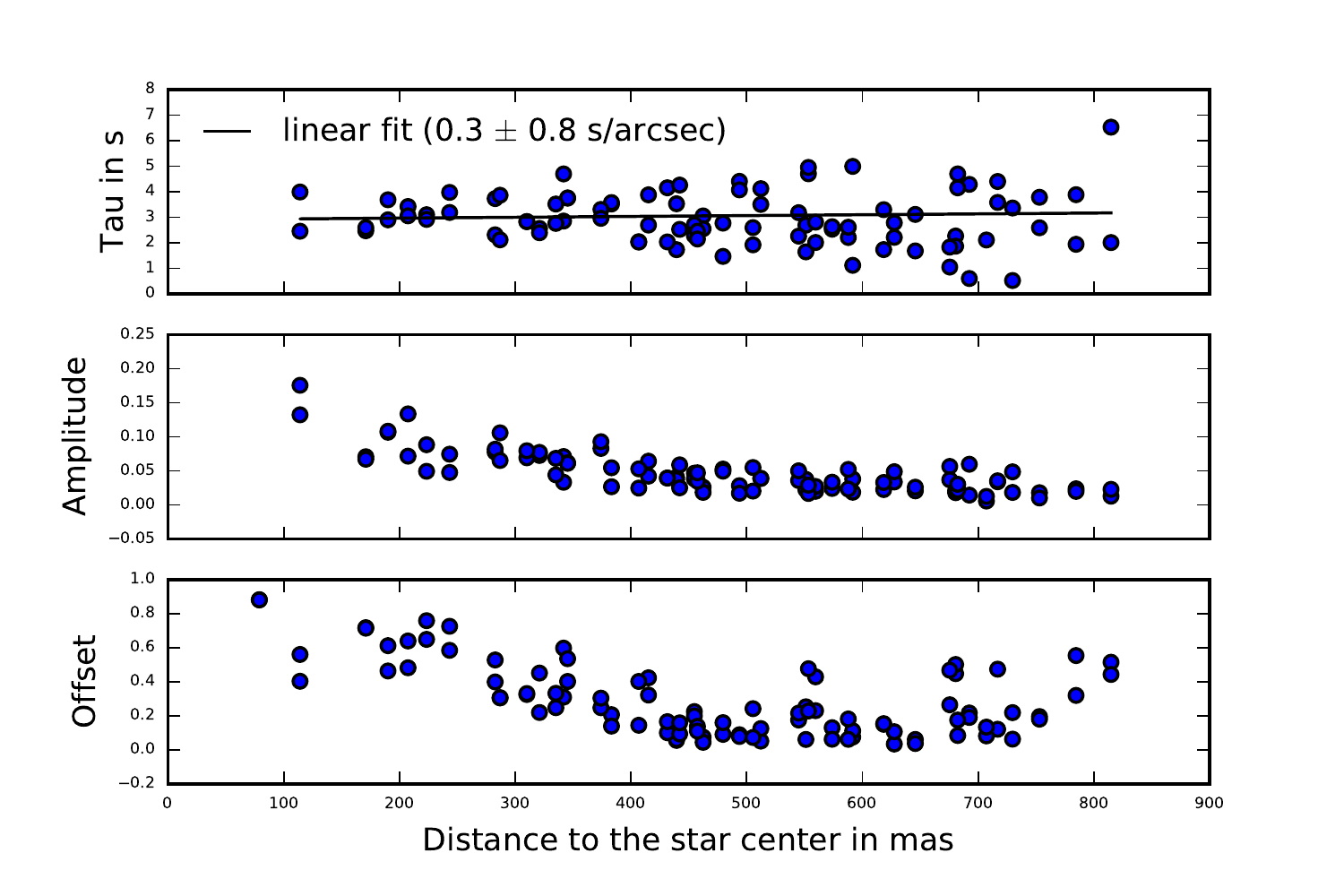}
	\end{tabular}
	\end{center}
   \caption[example] 
   { \label{fig_param_exp_fit_medsub} 
Parameters of the fit of an exponential decay (Eq. \ref{eq_exp_decay}) to the on-sky data. The top panel shows the characteristic time $\tau$ in s, the middle panel shows the amplitude $\Lambda$ and the bottom panel shows the offset $\rho_0$.}
   \end{figure}

In a third step, we analyzed the evolution of the correlation in the first minute. We can also notice in each subregion the fast decorrelation regime visible in the whole region \Stot{} on Fig. \ref{fig_corrcoeff_vs_time_expfit}. Therefore, we fitted for each subregion an exponential decay law, parametrized by Eq. \ref{eq_exp_decay}. The parameters of the fit are plotted in Fig. \ref{fig_param_exp_fit_medsub} as a function of the separation between the star and the center of the subregion \Ssub. 
In this case, we conclude that the characterstic time $\tau$ does not depend significantly on the separation, and is about 3\,ms (with slightly more scatter at larger separations because the S/N on speckles is smaller). A linear fit (black line in Fig. \ref{fig_param_exp_fit_medsub} top) is indeed compatible with a zero slope. A similar characteristic time $\tau$ was also found for the whole region \Stot, meaning that there is no zonal effect for the time it takes for the correlation to drop and reach its value after one minute. On the other hand, the amplitude and offset of the exponential law do significantly vary with the separation, which is easily explained because the correlations start with higher values close to the optical axis, as already stated. The same effect is also visible in the intercept of the linear regime (bottom plot of Fig. \ref{fig_param_linear_fit_vs_sep}).

\subsection{Comparison with the data obtained with the internal lamp}
\label{sec_comp_internal_data}

To disentangle the effects of atmospheric perturbations from the aberrations internal to the instrument, we performed the same analysis with an internal light source, as explained in section \ref{sec_measurements}. In this case, the only moving part in the instrument was the derotator which simulated pupil tracking of a star observed at an hour angle between -20 and +40\,min.  We performed the same data analysis as done for the on-sky data, namely a calculation of the correlation coefficient in the whole region \Stot{} and in subregions \Ssub.

\begin{figure} [ht]
   \begin{center}
   \begin{tabular}{c} 
   \includegraphics[width=\hsize]{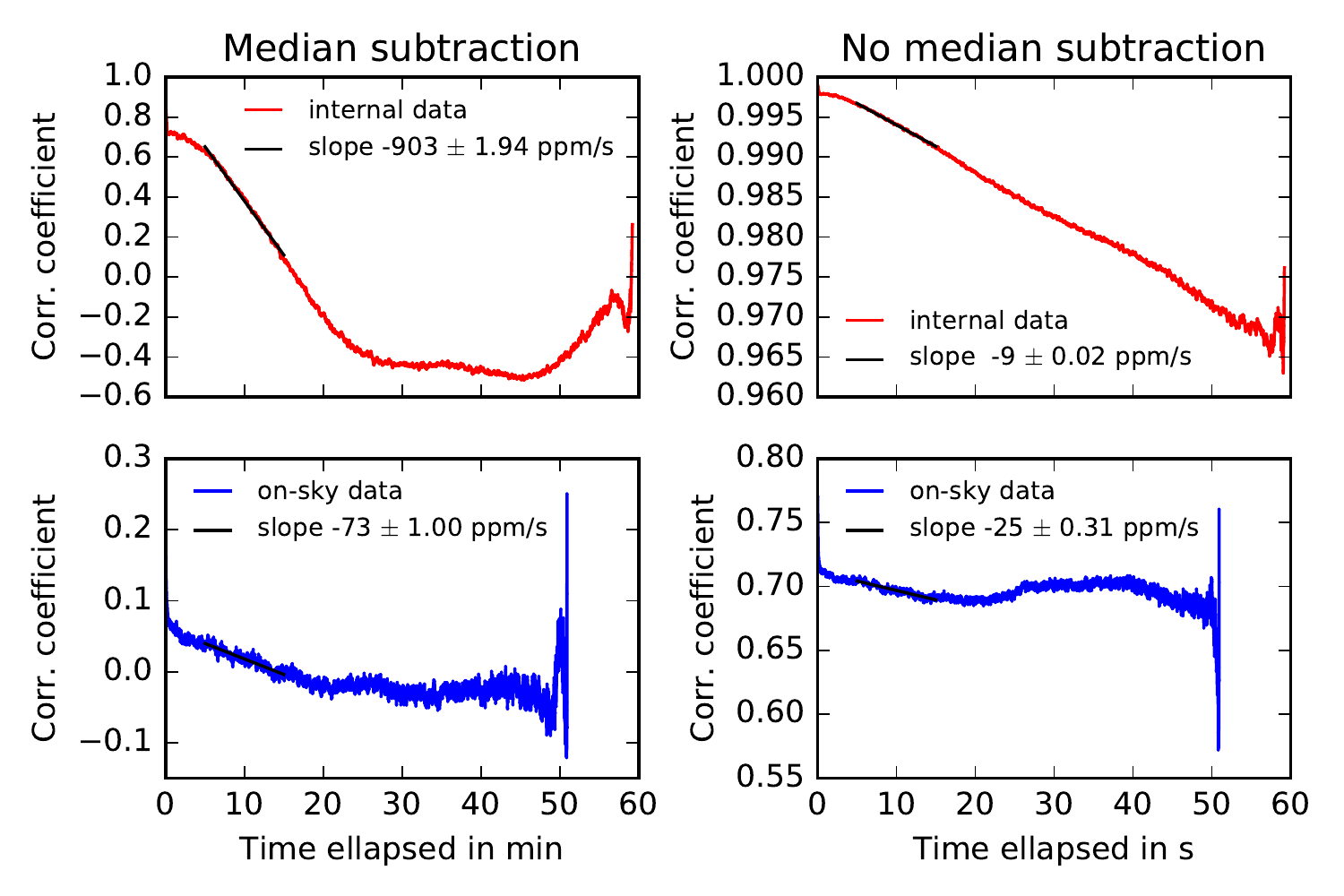}
	\end{tabular}
	\end{center}
   \caption[example] 
   { \label{fig_comparison_onsky_internal} 
Comparison between the decorrelation measured on-sky (blue points, identical as in Figure \ref{fig_corrcoeff_vs_time_linearfit} for the right panel) and the internal source (red points). Linear fits (black lines) was performed between 5 and 15\,min, the slope is indicated in the label. The correlation coefficients of the left panel were computed for a sequence of frames subtracted by the median frame, while those on the right panel were computed for the non-median-subtracted sequence.}
   \end{figure}

The comparison of the decorrelation curves with the on-sky data for the whole region \Stot{} is shown in Fig. \ref{fig_comparison_onsky_internal}. 
The more meaningful comparison of the two data sets can be done for sequence that has not been subtracted by the temporal median (right panels). Otherwise, the changes reflect the evolution of the residuals after subtraction of the median, which in the case of the internal data, are relatively low since the sequence is very stable. We however provide also the correlation coefficients in the median subtracted sequence for completeness. 

\begin{figure} [ht]
   \begin{center}
   \begin{tabular}{c} 
   \includegraphics[width=0.8\hsize]{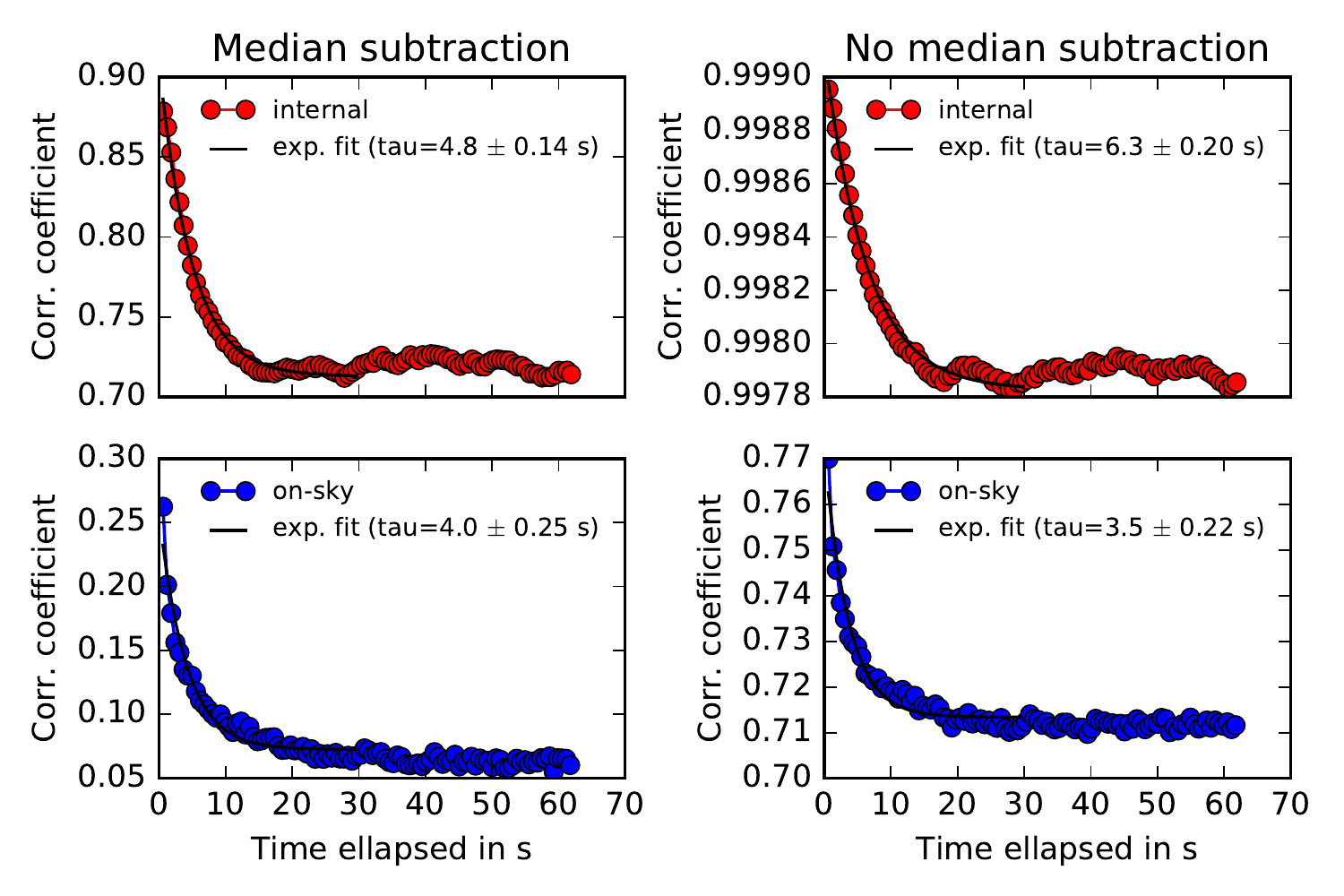}
	\end{tabular}
	\end{center}
   \caption[example] 
   { \label{fig_exp_fit_internal} 
Zoom of Fig. \ref{fig_comparison_onsky_internal} showing the evolution of the correlation coefficients over the first 60 seconds. A fit of an exponential profile was performed from 1 to 30\,s. The analysis was done on the internal and on-sky sequence with the temporal median removed (left plots) or not (right plots).}
   \end{figure}

As expected, because the sequence is very stable in time without any perturbation from the atmosphere, the correlation is higher, and stays above 95\% during one hour. We can define the same two regimes of decorrelation: a linear regime extending all the way from a few minutes to one hour, and, surprisingly, a steep decorrelation during the first seconds, shown in Fig. \ref{fig_exp_fit_internal} .
As expected, the linear speed of decorrelation (computed for both internal and on-sky data between 5 and 15min) is smaller for the internal data by almost a factor three, indicating that the PSF decorrelates more slowly in the absence of atmospheric perturbations. Surprisingly, the same steep decorrelation occurs on a time scale of a few seconds. The characteristic time $\tau$ is slightly larger for the internal data than for the on-sky data ($6.3\pm0.2$s vs $3.5\pm0.2$s, right panel of Fig. \ref{fig_exp_fit_internal}) but this is is less obvious once the temporal median of the sequence has been removed to enhance the relative variation of the quasi-static speckles disregarding the static part of the PSF ($4.8\pm0.2$s vs $4.0\pm0.3$s, left panel of Fig. \ref{fig_exp_fit_internal}).


\begin{figure} [ht]
   \begin{center}
   \begin{tabular}{c} 
   \includegraphics[width=0.5\hsize]{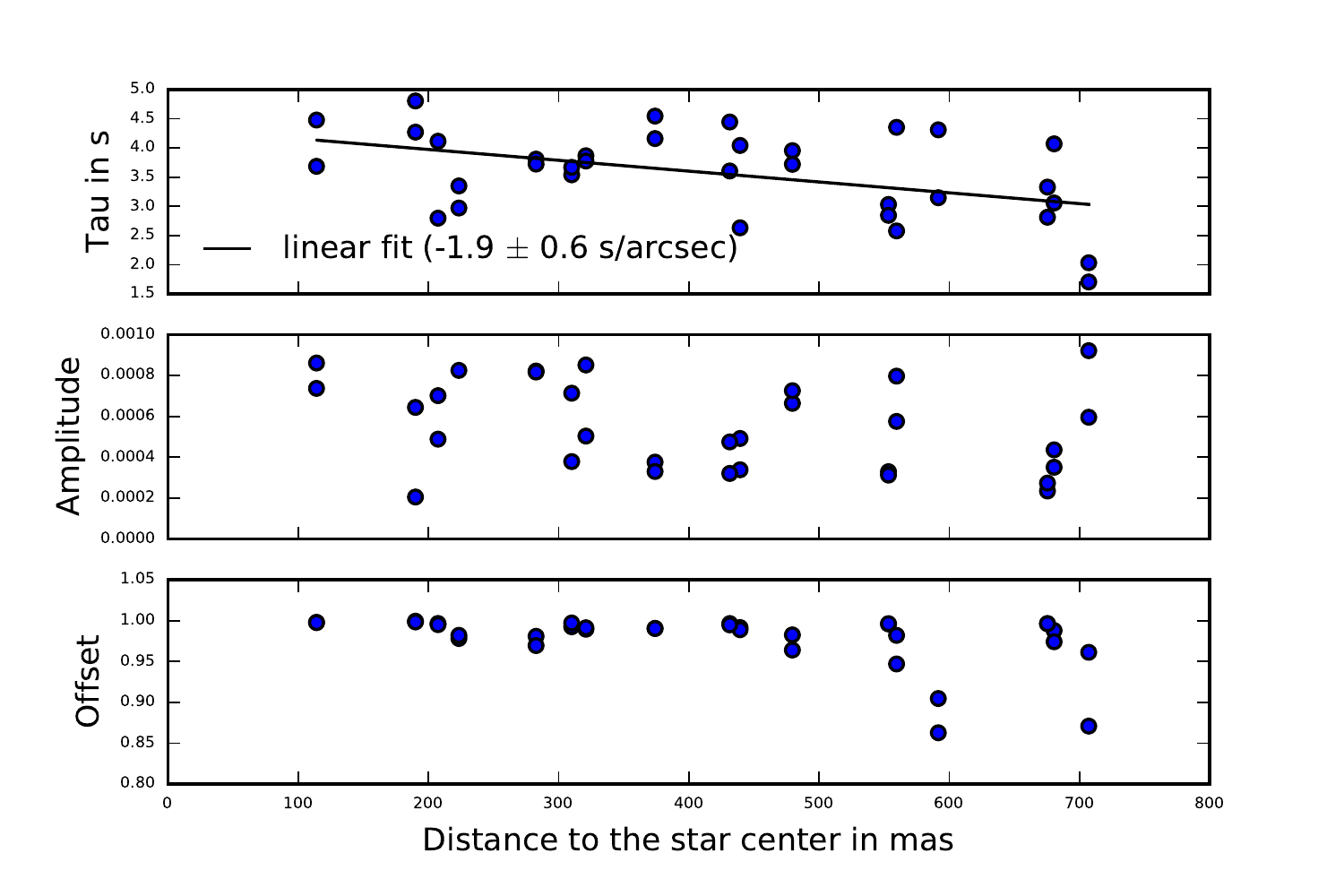}
   \includegraphics[width=0.5\hsize]{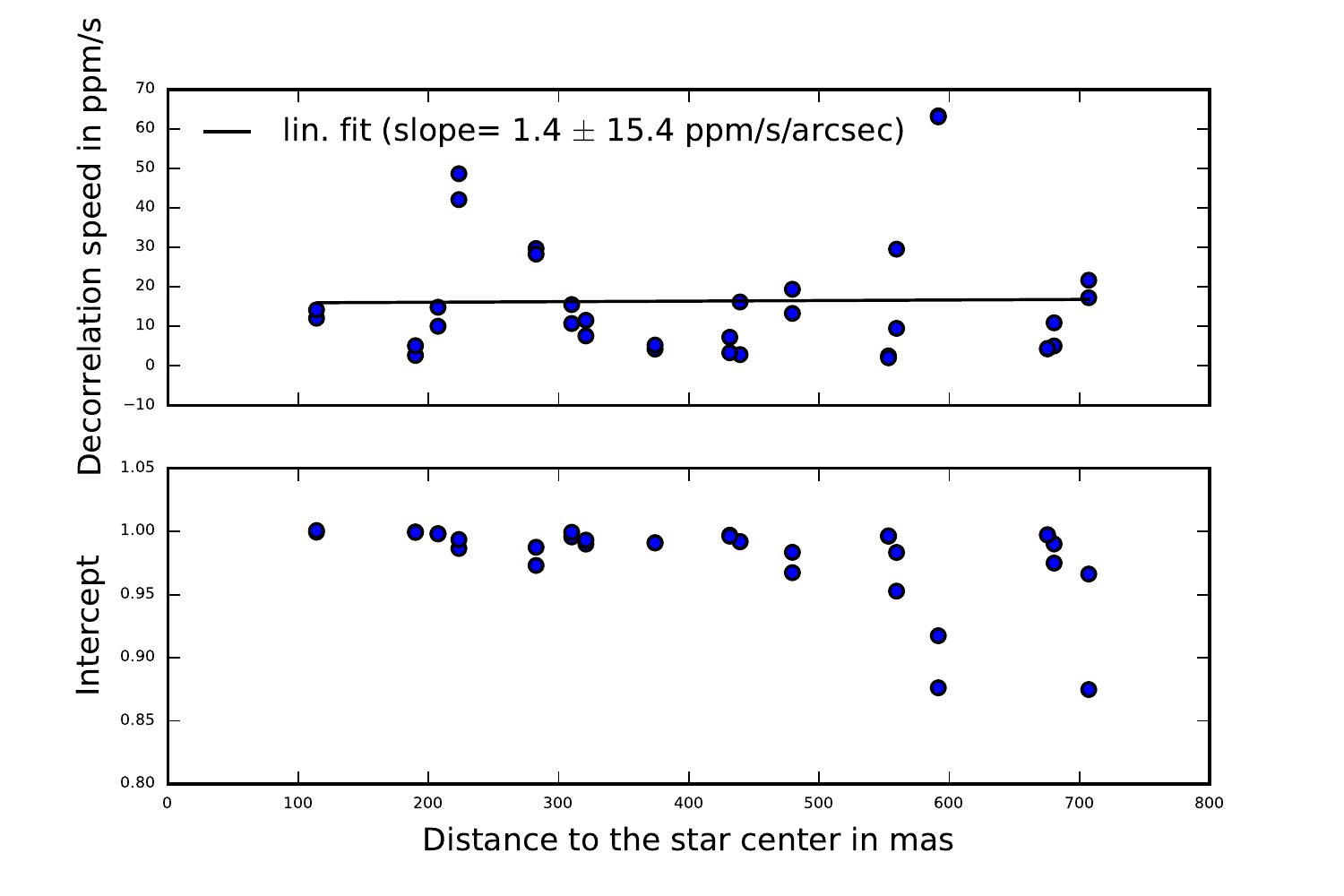}
	\end{tabular}
	\end{center}
   \caption[example] 
   { \label{fig_param_exp_fit_internal} 
  Parameters of the fit of an exponential decay (left panel, as parametrized by Eq. \ref{eq_exp_decay}) and of the linear fit (right panel) to the internal data in each subregion \Ssub{} after the temporal median subtraction. Each data point correspond to an individual subregion. The top left panel shows the characteristic time $\tau$ in s, the middle left panel shows the amplitude $\Lambda$ and the bottom left panel shows the offset $\rho_0$. The top right panel represents the decorrelation speed  (opposite of the slope) and the bottom right panel the intercept.}
   \end{figure}

A zonal analysis of the internal data shows that the characteristic time of the steep exponential decay in the first seconds is only marginally dependent on the separation to the central source, as visible in Fig. \ref{fig_param_exp_fit_internal} left (the black line shows a linear fit only marginally decreasing with the separation, within error bars).

\section{Interpretations and additional tests}

We will here review some possible explanations for the behaviours presented in section \ref{sec_results}.

\subsection{A linear decorrelation with time}

Between 5 and 20\,min, the correlation decreases in a linear way. The atmospheric parameters on such a timescale were stable, we can therefore reasonably rule out  external perturbations as the origin of this trend. This regime is also seen with the internal source, although the decorrelation happens at a slower rate. This confirms that the source of this regime is internal to the instrument, and also unrelated to the mechanical flexures of the telescope mirrors. Moreover, we carried out a control experiment by computing the decorrelation in an empty sky region of the detector, and we confirmed the absence of such a linear decrease in this case, detailed in Appendix \ref{app_sky}. We showed that this linear regime evolves faster at short separations, as visible in Fig. \ref{fig_param_linear_fit_vs_sep}. This means that the low spatial frequency aberrations (which correspond to regions at small separations in the focal plane) evolve faster than the small scale aberrations in the pupil (corresponding to regions at larger spearations in a focal plane). An additional test carried out with the internal lamp with the AO loop closed but all components fixed in the instrument (including the derotator, that was moving in the measurements presented in section \ref{sec_comp_internal_data}), shows in this case that the decorrelation speeds are scattered about a zero mean, for the different subregions \Ssub. This tends to show that this slow linear regime is related to the moving components inside the instrument. A contribution from the thermal expansion of these components might also be involved and could explain why we observe an evolution of small spatial frequencies: the thermal evolution inside the instrument enclosure is of the order of several tens of minutes and induce large-scale aberrations on the optics. 

If this explanation holds true, this linear regime should evolve slower for an observation carried out about meridian and faster far away after or before meridian. Indeed, for a Nasmith instrument such as VLT / SPHERE, in pupil-stabilized mode, the derotator speed cancels at meridian passage and is large far way from meridian. We used therefore our on-sky data set to test this hypothesis, but could not come to a clear-cut conclusion on this question. The analysis is shown in Appendix \ref{sec_influence_ha}. Our observations were performed after meridian passage, and to study this effect specifically, longer observations about meridian in stable conditions should be repeated. A similar analysis was done with VLT / NaCo around meridian for a longer temporal sequence and could confirm this trend\cite{Milli2016_HIRES}. 

\subsection{A steep decorrelation over a few seconds}

A steep regime of decorrelation occurs in the first 30\,s, well explained by an exponential decay with a characteristic time of about four seconds. For such a short time scale, ruling out the atmospheric turbulence requires more analysis. Indeed the coherence time was about 3.5\,ms in the optical (as estimated from a combination of DIMM measurements and wind speed predictions at 200mbar\cite{Sarazin2002}), which can be converted to a value of 15\,ms in the H band. This is too small to explain the steep decorrelation by the presence of atmospherical perturbations uncorrected by the AO system. Macintosh et al. \cite{Macintosh2005} showed that residual speckles caused by non-corrected atmospheric perturbations are refreshed on the pupil every $D/\nu$, where $D$ is the telescope diameter and $\nu$ the wind speed, inducing a typical decorrelation time of $0.6 D/\nu$. The wind speed during the observations was 3 to 4 m/s on the ground and higher in altitude, meaning that the resulting time scale is $0.6 D/\nu\leq1.6$s. This is about the same order of magnitude of what is measured on-sky. However, this fast decorrelation regime is also seen with the internal lamp. It is therefore not an effect of the atmospheric perturbations, or the non corrected atmospheric residual phase screen refreshed every second. 
Moreover, all spatial scales evolve at the same speed because the characteristic times inferred for each individual region \Ssub{} do not depend on the separation to the central source. We can also safely rule out any detector persistance effect, as shown in Appendix \ref{app_remanence}.
In an additional attempt to characterize this effect, we investigated if this can be related to the temporal sampling of the sequence. Frames were recorded every  0.6\,s, which is close to the time scale revealed here. We recorded at a later epoch a sequence of frames every 0.18\,s and describe the analysis in Appendix \ref{app_sampling}. This 4\,s timescale is still visible (Fig. \ref{fig_exp_fit_onsky_high_cadence}), despite much different atmospheric conditions, in particular high wind. We can therefore conclude from these analysis that this fast decorrelation over a few seconds is real, related to an internal effect in the intrument independent of the atmospheric conditions or the telescope. 

\section{CONCLUSION}

We studied the decorrelation timescale of a deep sequence of SPHERE coronagraphic images on a bright star. This sequence is unique because we recorded frames at a high rate of 1.6\,Hz by windowing the SPHERE / IRDIS camera in stable atmospheric conditions with a very good AO correction. We used the Pearson correlation coefficients between pairs of frames to evaluate the relative change in the frame speckle pattern during the time ellapsed between the two frames acquisitions. Our analysis highlights two distinct regimes of decorrelation. On the one hand, we observe a linear decorrelation happening for ellapsed times between a few minutes and 20\,min. This regime is also observed with an internal lamp, the instrument shutter closed and the derotator simulating pupil-tracking, showing that this effect is likely related to the mechanical movements and thermal expansion within the instrument. Additional tests are needed to confirm the origin of this regime, especially a sensitivity analysis with the hour angle, to show that the decorrelation speed is smaller during meridian passage. 
On the other hand, we also observe a steeper regime of decorrelation taking place during the first few seconds and well fit by an exponential decay. Again this effect is also seen with the internal lamp, shutter closed, showing that this is also an effect internal to the instrument, although the exact cause of this decorrelation is not clear yet.  

\appendix   

\section{Evolution of the correlation coefficient with time in the subregions}
\label{app_subregions}

\begin{figure} [ht]
   \begin{center}
   \begin{tabular}{c} 
   \includegraphics[width=\textwidth]{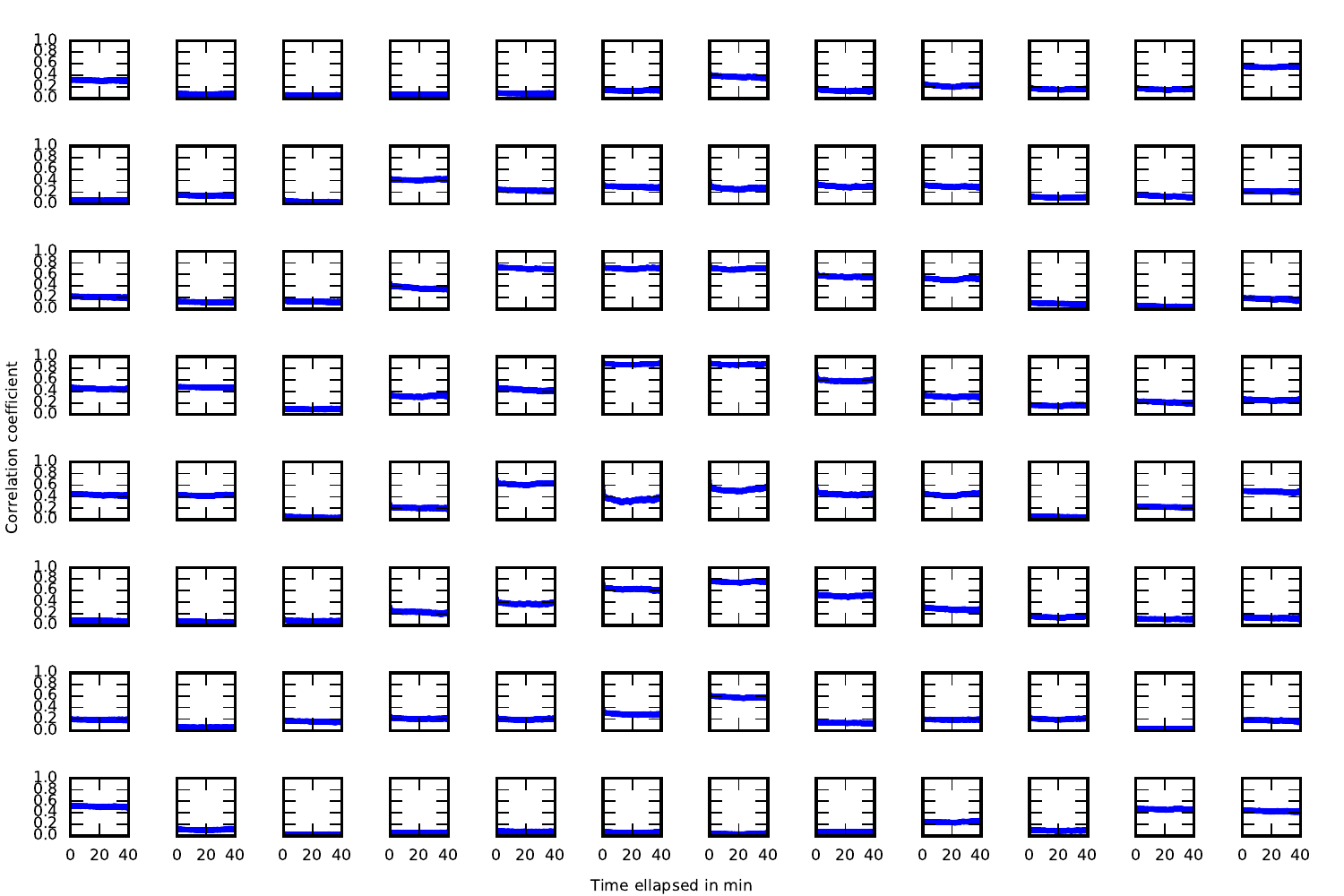}
	\end{tabular}
	\end{center}
   \caption[example] 
   { \label{fig_onsky_bigpicture} 
Plot the correlation coefficients vs time for 8 by 12 subregions (out of the 12 by 12). The position of each curve in the panel corresponds to the position of the area \Ssub where the correlation is computed, as shown in Fig. \ref{fig_grid_subregions} }
   \end{figure}

Fig. \ref{fig_onsky_bigpicture} shows a global image of 8 by 12 correlation coefficient curves for the zonal analysis in the subregions \Ssub. The correlation is higher in the center, close to the coronagraph, as also shown by Fig. \ref{fig_param_exp_fit_internal}. 

\section{VERIFICATIONS}

\subsection{Analysis of the correlation in an empty sky region}
\label{app_sky}

\begin{figure} [ht]
   \begin{center}
   \begin{tabular}{c} 
    \includegraphics[width=0.8\hsize]{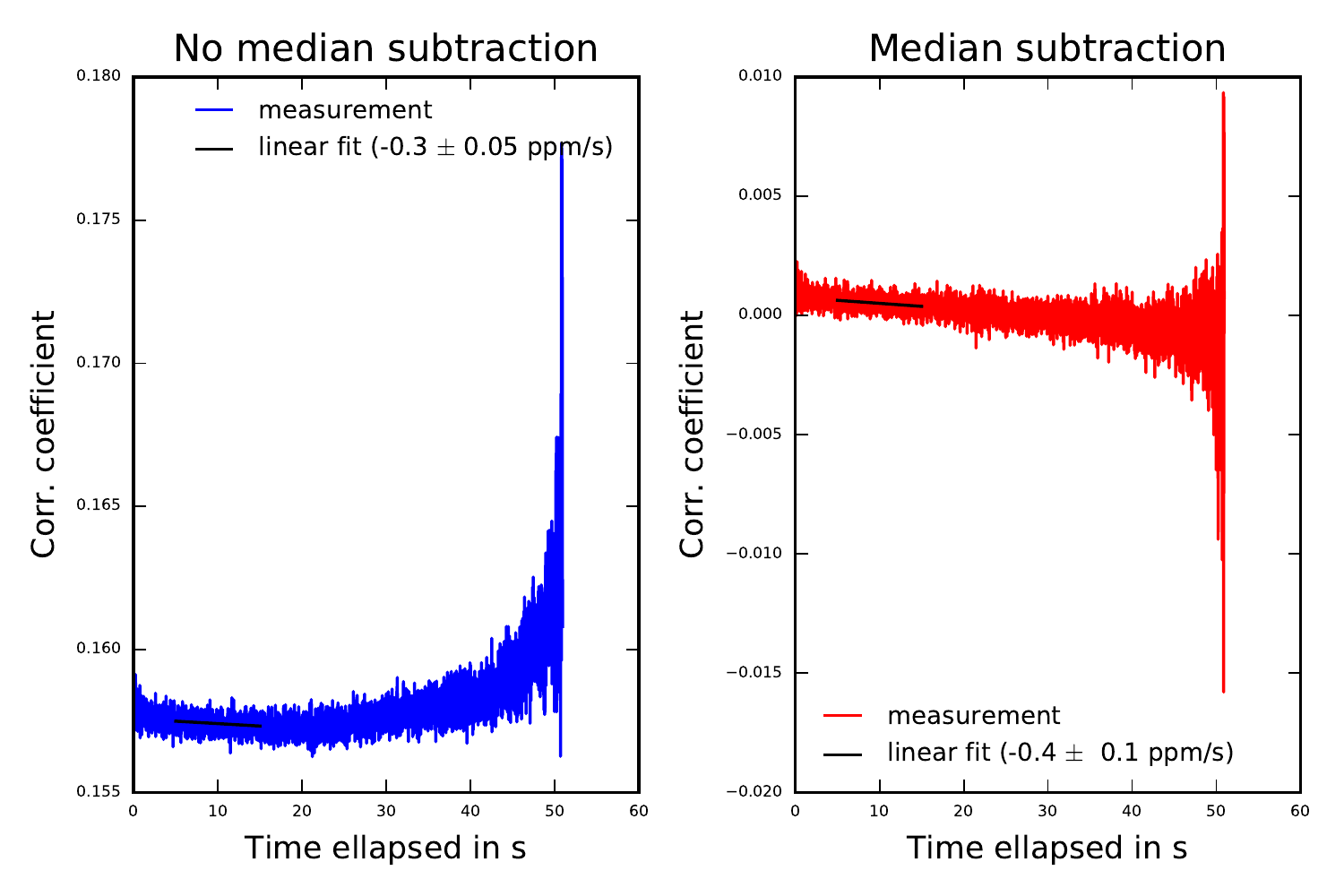}
	\end{tabular}
	\end{center}
   \caption[example] 
   { \label{fig_corr_coeff_sky} 
Control experiment: correlation coefficient versus time computed in a sequence of an empty sky region, with as many pixels as in the baseline region \Stot. The left panel shows the result on a sequence with a non-zero temporal mean, and the right panel shows the results once the temporal median has been subtracted.}
 \end{figure}

To verify the validity of our measurement procedure, we proceeded to various control experiments. On of them consists in applying the same procedure to a sequence of frames containing only background and thermal noise, without any stellar signal. This was done by using a region of the 128 px wide stripe read on the detector at several arcseconds from the central star, with the same size as the region \Stot. The result is shown in Fig. \ref{fig_corr_coeff_sky}. The mean correlation is 0.157 between 5 and 15min and a linear fit to the data yields a zero slope, confirming that the decorrelation seen in the region well-corrected by the AO system is related to the starlight, its propagation in the atmosphere and in the instrument and its correction by the AO system. This analysis also enables us to get the uncertainty associated to the estimation of the correlation coefficient, measured as the dispersion of the values. It increases towards larger ellapsed time because the number of pairs of frames separated by $\Delta t$ decreases as $\Delta t$ increases.

\clearpage

\subsection{Remanence of the detector}
\label{app_remanence}

\begin{figure} [ht]
   \begin{center}
   \begin{tabular}{c} 
 \includegraphics[width=0.5\hsize]{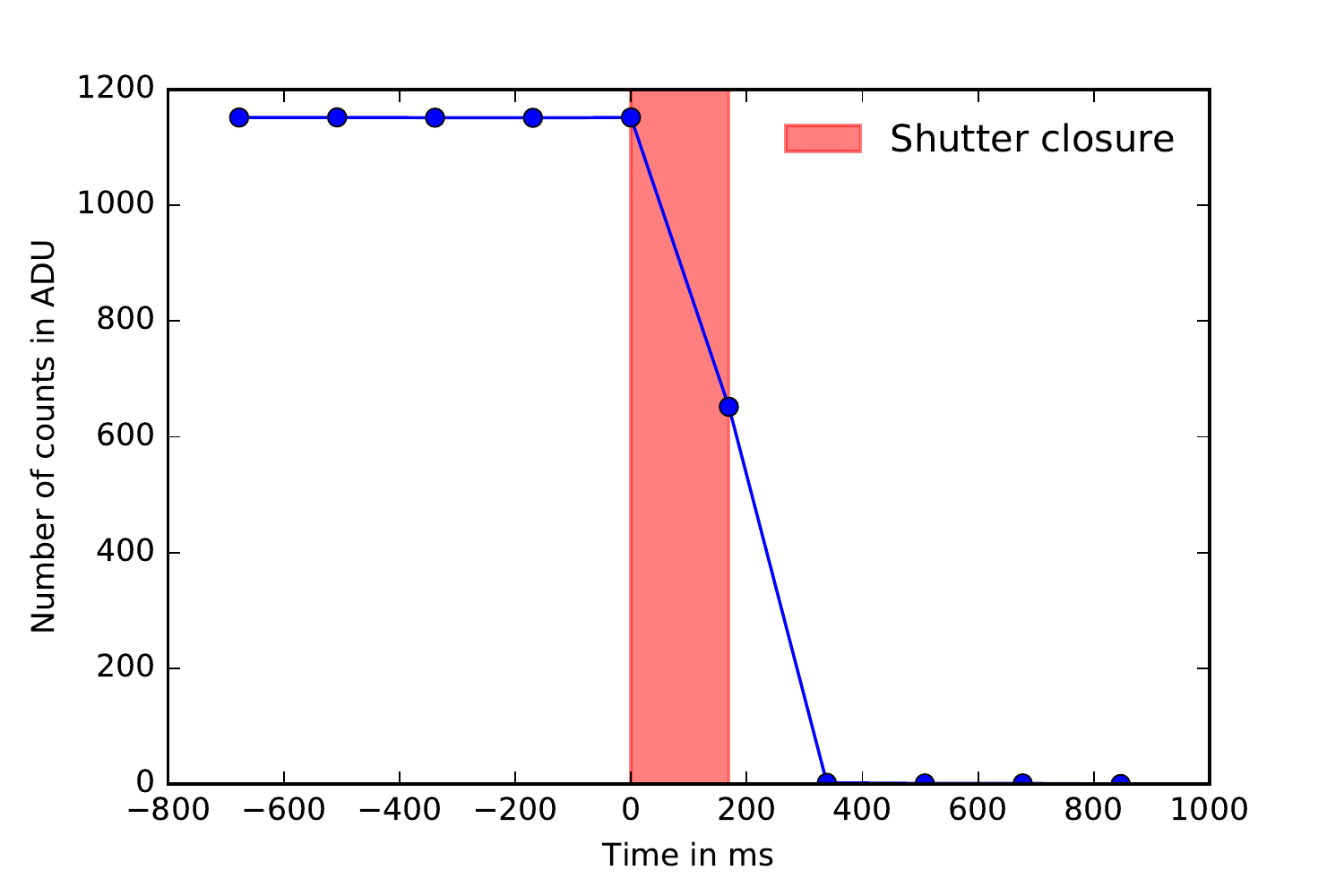}
	\end{tabular}
	\end{center}
   \caption[example] 
   { \label{fig_shutter_closure} 
Evolution of the mean number of counts in a 128 columns by 20 rows stripe of the detector illuminated by a flat lamp. Each detector read (blue circle) is separated by 169ms. The SPHERE/IRDIS shutter was closed in the red-shaded region.
}
 \end{figure}

The SPHERE / IRDIS detector is subject to remanence when a strong illumination is applied. Although the detector was not saturated during the tests presented in this analysis, we verified that the fast decorrelation seen over a few seconds could be due to remanence effects in a region strongly illuminated just outside the coronagraph. To do so, we recorded a sequence of images with a uniform illumination of the detector, e.g. a flat field sequence, and closed the IRDIS shutter during the sequence to study the decay of the detector counts. We windowed the detector to read only 20 rows and be able to have a detector integration time (DIT) of only 25ms. Due to the readout time, we could record a frame every 169ms. For such a short DIT, the illumination was about 1150\,ADU, about ten times higher than the level recorded in the baseline test scenario presented here. We computed the mean number of counts in the detector stripe, and show it in Fig. \ref{fig_shutter_closure} in a window centered on the shutter closure (red-shaded area). The count level drops to zero within to readouts, meaning that we can rule out any persistence effect in our analysis for a timescale greater than 340ms. 

\subsection{Effect of the time sampling}
\label{app_sampling}

\begin{figure} [ht]
   \begin{center}
   \begin{tabular}{c} 
 \includegraphics[width=0.6\hsize]{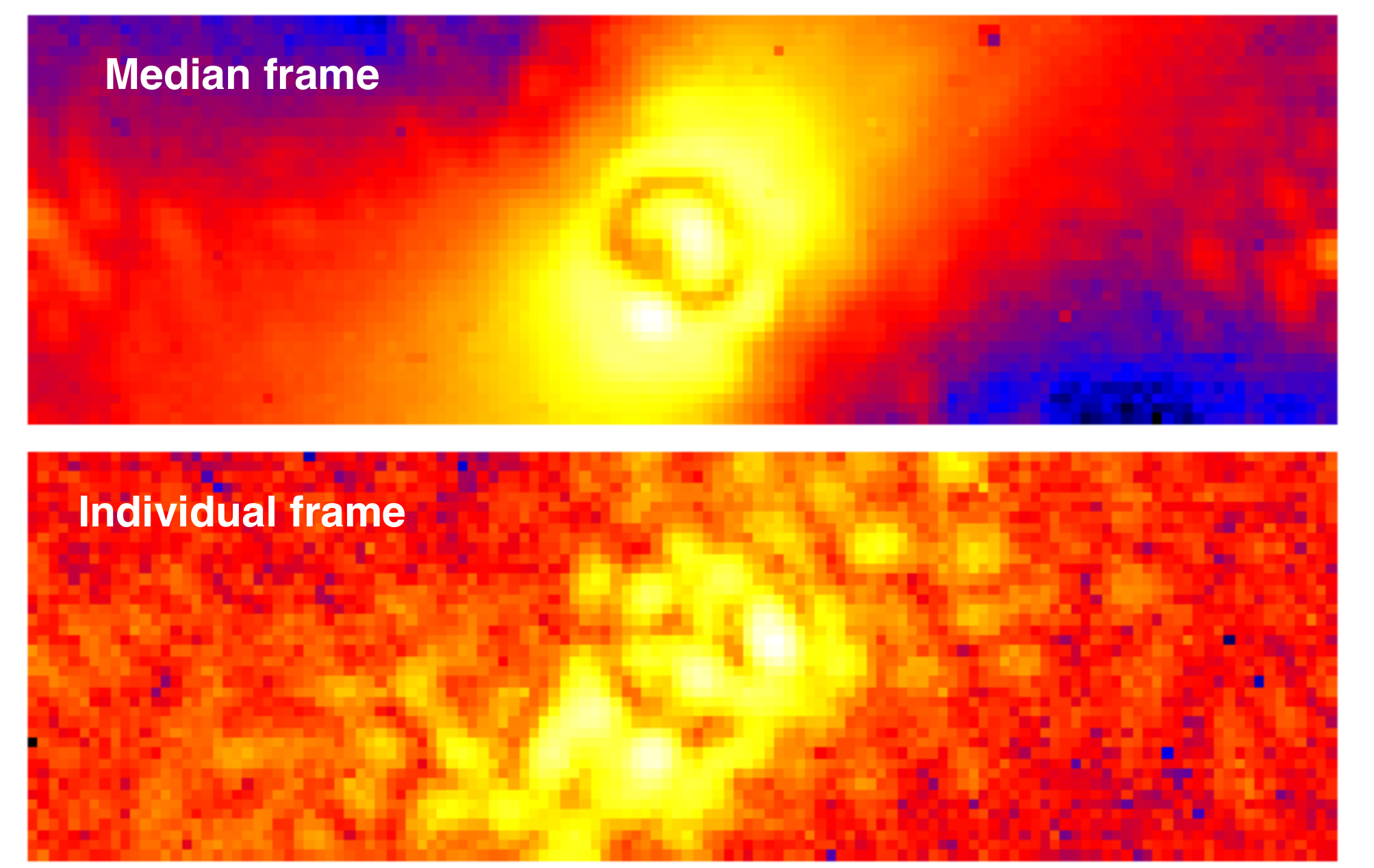}
	\end{tabular}
	\end{center}
   \caption[example] 
   { \label{fig_median_onsky_high_cadence} 
Images recorded on June 4 2016 with a high frame rate of 5.6\,Hz. The detector was windowed to 1.57\arcsec by 0.49\arcsec. Top: median of the sequence, bottom: example of a individual DIT of 52\,ms. The median image appears blurry because the PSF elongation is due to the high wind and its direction follows the parallactic angle.}
 \end{figure}

\begin{figure} [ht]
   \begin{center}
   \begin{tabular}{c} 
    \includegraphics[width=0.8\hsize]{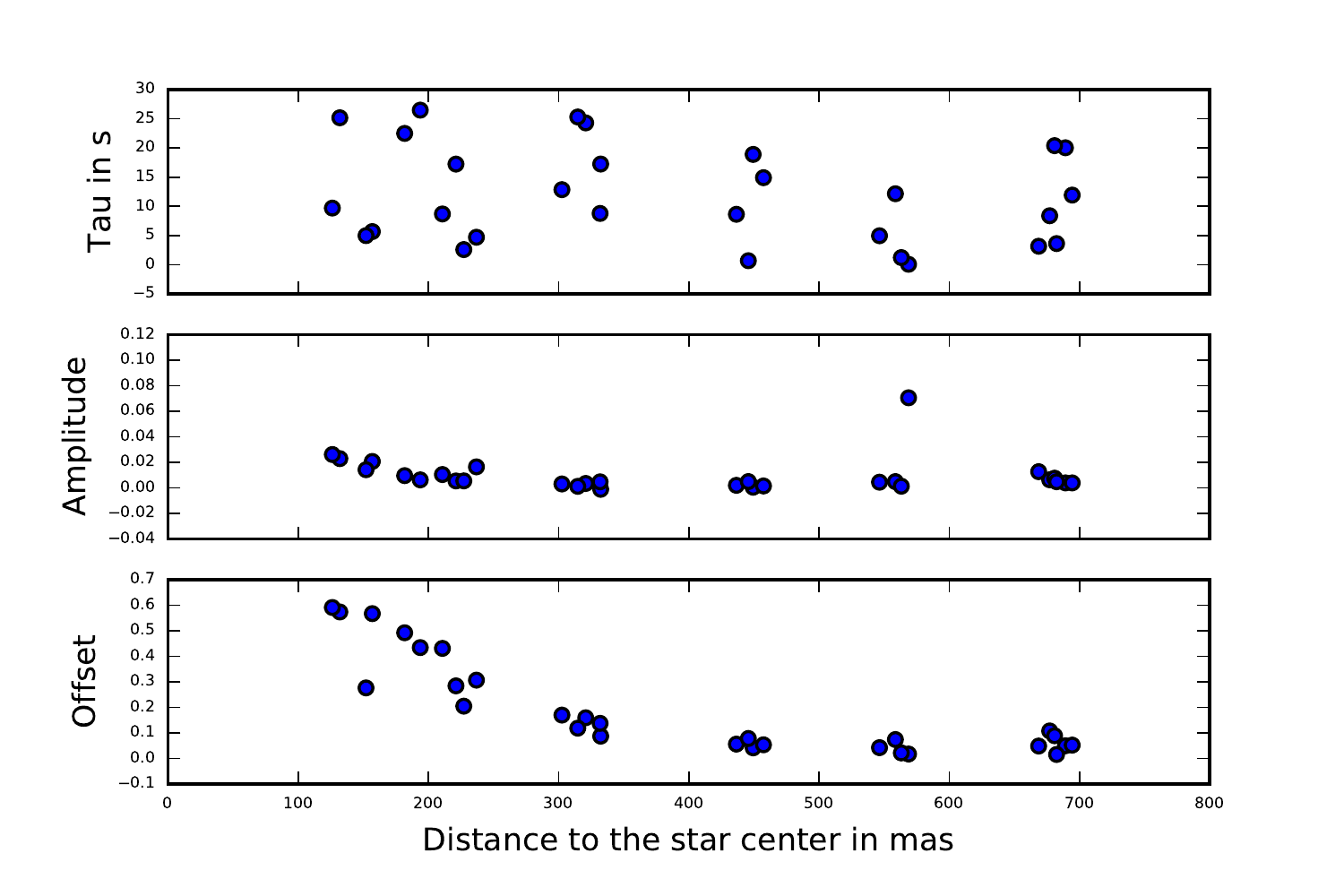}
	\end{tabular}
	\end{center}
   \caption[example] 
   { \label{fig_exp_fit_onsky_high_cadence} 
 Parameters of the fit of an exponential decay to on-sky data taken with a high cadence of 176ms or 5.6Hz. The top panel shows the characteristic time $\tau$ in s, the middle panel shows the amplitude $\Lambda$ and the bottom panel shows the offset $\rho_0$.}
 \end{figure}

In section \ref{sec_decorr_regime} and \ref{sec_exp_fit_vs_sep}, we show that a fast decorrelation of the images occur in a timescale $\tau$ of about 4s, both for the whole region \Stot{} and in each subregion \Ssub. The sampling frequency of 1.6\,Hz is relatively close to the time scale, and the first data point in Fig. \ref{fig_corrcoeff_vs_time_expfit} is higher than the best fit value (red curve), therefore we investigated whether a faster image frame rate would lead to a different estimate of $\tau$. To do so, we windowed the detector to read only 40 raws (equivalent to 0.49\arcsec on sky), as visible in Fig. \ref{fig_median_onsky_high_cadence}. This way, we could get a detector integration time (DIT) of 52\,ms, and record a frame every 177\,ms, equivalent to a frame rate of 5.6\,Hz. This sequence was recorded under poor conditions (coherence time of 2.4ms, DIMM seeing 0.7\arcsec but very high altitude wind and high wind speed on the ground of 12m/s) on the star $\beta$ Crux of magnitude R=1.4, on the night of June, 4 2016. Despite those adverse wind conditions, the Strehl was 80\%, but a significant elongation of the PSF is visible along the diagonal of the detector, and induced significant light leak outside the coronagraph radius.
We computed the correlation in subregions \Ssub with the same size as in our baseline sequence. The parameters derived from a fit of an exponential decay (Eq. \ref{eq_exp_decay}) are shown in Fig. \ref{fig_exp_fit_onsky_high_cadence}. The values of the characteristic time $\tau$ are spread about 4\,ms, a value almost identical as that found for the baseline sequence sampled at 1.6\, Hz, with some scatter at larger separation where the S/N on the speckles is lower. Therefore we can conclude that the timescale of about 4\,s derived for the fast decorrelation regime is robust and does not depend on the frame rate at which the sequence is acquired. Shorter DITs on a brighter star lead to the same estimate of $\tau$. In addition, this additional observation also shows that $\tau$ does not seem to be related to the quality of the AO correction or the atmospheric conditions, because the Strehl was in this case more than 10\% lower than the baseline scenario and the strong wind induced a low coherence time.

\section{Link with the contrast}
\label{app_contrast}

The decorrelation speeds might seem small, nevertheless, the contrast which is the relevent quantity that one aims at minimizing does not evolve linearily with the correlation. For a pair of images $(I_i,I_j)$ reduced and centered (e.g subtracted by the spatial mean in the region \S{} and divided by the standard deviation over \textsl{S}), the instantaneous contrast as defined by  the $\mathcal{L}^2$ norm of the simple difference $I_i-I_j$ can be expressed as
\begin{equation}
C=\sqrt{2(1-\rho)}
\end{equation}
where $\rho$ is the correlation coefficient as defined in Eq. \ref{eq_def_corr_coeff}. In such a case, the rate of change of the contrast with time $\frac{dC}{dt}$ is related to the correlation by
\begin{equation}
\frac{dC}{dt} = -\frac{1}{C} \frac{d\rho}{dt}
\end{equation}
This means that for an instantaneous contrast $C=10^{-3}$ and a decorrelation speed $\frac{d\rho}{dt}=73 \text{ppm/s}$, the contrast decreases at a very high speed of 73\,000 ppm/s.

\section{INFLUENCE OF THE HOUR ANGLE ON THE DECORRELATION}
\label{sec_influence_ha}

\begin{figure} [ht]
   \begin{center}
   \begin{tabular}{c} 
   \includegraphics[width=0.8\hsize]{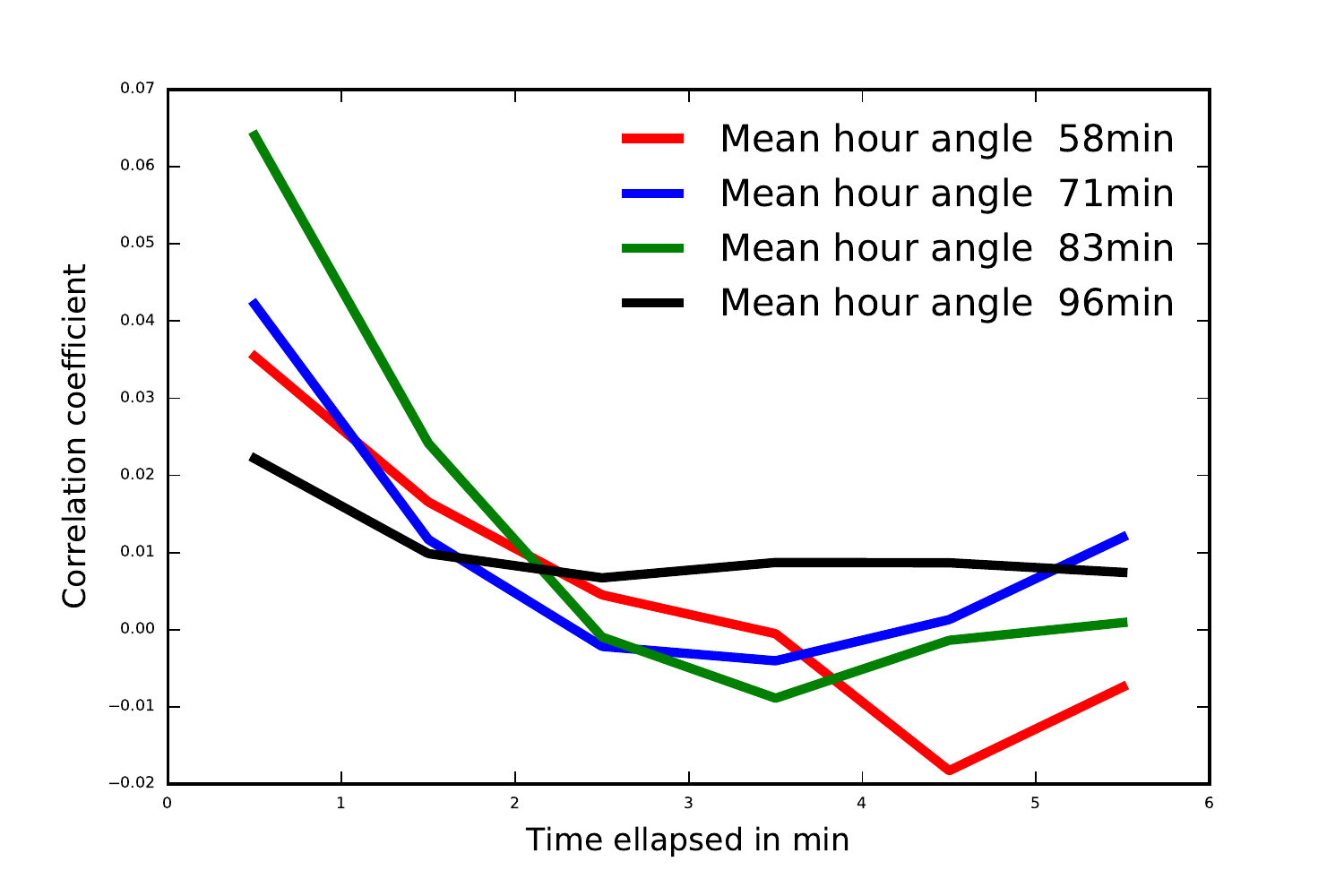}
	\end{tabular}
	\end{center}
   \caption[example] 
   { \label{fig_influence_ha} 
Correlation coefficient versus time for 4 batches of frames grouped according to their distance to the meridian, e.g. the hour angle. The data have been binned every minute to increase the S/N on the correlation coefficient.}
   \end{figure}

For a telescope with an alt/az mount observing in pupil tracking mode, the quasi-static speckles are expected to be the most stable at meridian passage. At this specific moment, the tracking speed of the altitude axis is indeed going to zero, meaning that the aberrations of the optical parts induced by mechanical flexures of the telescope are minimzed. In addition, the derotator speed, which for a Nasmyth instrument follows the altitude axis, also is moving at its smallest derotation speed, and the ADC movements are minimzed because the target is the highest in the sky. For all these reasons, we tried to detect any influence of the hour angle in our analysis, as it was done for VLT/NaCo data\cite{Milli2016_HIRES}. We therefore binned the on-sky data of 3087s into four batches of 13min each and calculated the evolution of the correlation coefficient over time for each batch. The result is shown in Fig. \ref{fig_influence_ha}. It does not show any significant influence of the hour angle. However, the star was observed after meridian passage, therefore the analysis should be repeated on a sequence centered around meridian to get a higher sensitivity to influence of the hour angle. In addition it should be highlighted that this kind of sensitivity analysis requires a very stable sequence of observations to be able to make a meaningful comparison of the different batches and rule out any effect of the evolution of the atmospheric conditions between the batches.  

\acknowledgments 
 
 J.M. acknowledges support from the ESO fellowship program and we thank ESO staff and technical operators at the Paranal Observatory.

\bibliography{report} 
\bibliographystyle{spiebib} 

\end{document}

%% file: macros_spie.tex
\newcommand {\micron}{\mbox{$\mu$m}}
\newcommand\arcsec{$^{\prime\prime}$}